\documentclass{aa}

\usepackage{txfonts}

\usepackage{graphicx}   

\usepackage{lscape}
\newcommand{\Teff}{$T_{\rm eff}$}
\newcommand{\Fcl}{$F_{\rm cl}$}
\newcommand{\Vcl}{$V_{\rm cl}$}
\newcommand{\Vinf}{$V_{\rm \infty}$}
\newcommand{\logg}{log\,\textit{g}}
\newcommand{\vsini}{\textit{v}\,sin\,\textit{i}}

\newcommand{\Mdot}{\textit{\.M}}
\DeclareRobustCommand\ion[2]{%
  \mbox{#1\kern0.2em%
  \smaller\rmfamily%
  \edef\@tempa{\@car#2\@nil}%
  \ifcat1\@tempa%
  \@Roman{#2}%
  \else%
  \uppercase{#2}%
  \fi}}
\def\ion#1#2{\ensuremath{\mathrm{#1}\;}{\protect\small\rm{#2}}}
\DeclareRobustCommand\nodata{$\cdots$}
\newcommand{\Msun}{\textit{M}$_{\odot}$}
\begin{document}

\title{Creating and using large grids of precalculated model atmospheres for a rapid analysis
of stellar spectra}
\author{J. Zsarg\'o \inst{1,2} 
   \and
   C. R. Fierro-Santill\'an \inst{3}
   \and
   J. Klapp \inst{1,3} \email{jaime.klapp@inin.gob.mx}
   \and
   A. Arrieta \inst{4} 
   \and
   L. Arias \inst{4} 
   \and
   J. M. Valencia \inst{2}
   \and
   L. Di G. Sigalotti \inst{5}
   \and
   M. Hareter \inst{5} 
   \and
   R. E. Puebla \inst{6,7} 
}

\institute{
Environmental Physics Laboratory (EPHYSLAB), Facultad de Ciencias, Campus de Ourense,
Universidad de Vigo, Ourense 32004, Spain
\and
Departamento de F\'{i}sica, Escuela Superior de F\'{i}sica y Matem\'aticas,
Instituto Polit\'ecnico Nacional, Avenida Instituto Polit\'{e}cnico 
Nacional S/N, Edificio 9, Gustavo A. Madero, San Pedro Zacatenco, 
07738 Ciudad de M\'{e}xico, Mexico 
\and
Departamento de F\'{\i}sica, Instituto Nacional de Investigaciones Nucleares (ININ), 
Carretera M\'exico-Toluca km. 36.5,
La Marquesa, 52750 Ocoyoacac, Estado de M\'exico, Mexico,
\and
Universidad Iberoamericana, Prolongaci\'on Paseo de la Reforma 880,
Alvaro Obregon, Lomas de Santa Fe, 01219 Ciudad de M\'exico, Mexico,
\and
Departamento de Ciencias B\'asicas, Universidad Aut\'onoma Metropolitana-Azcapotzalco 
(UAM-A), Av. San Pablo 180, 02200 Ciudad de M\'exico, Mexico
\and
Facultad de Ingenier\'{i}a, Ciencias F\'{i}sicas y Matem\'atica, Universidad Central del 
Ecuador, Ciudadela Universitaria, Quito, Ecuador
\and
Centro de F\'{i}sica, Universidad Central del Ecuador, Ciudadela Universitaria, Quito, Ecuador
}

\date{Received MM DD, 2020; Accepted MM DD, 2020}

\label{firstpage}

\abstract{}
              { We present a database of 43,340 atmospheric models ($\sim$80,000 models at the 
                conclusion of the project) for stars with stellar masses between 9 and 120 $M_{\odot}$, 
                covering the region of the OB main-sequence and Wolf-Rayet (W-R) stars in the 
                Hertzsprung--Russell (H--R) diagram.}
              { The models were calculated using the ABACUS I supercomputer and the stellar
                atmosphere code CMFGEN.}
              { The parameter space has six dimensions: the effective temperature \Teff, the 
                luminosity $L$, the metallicity $Z$, and three stellar wind parameters:
                the exponent $\beta$, the terminal velocity $V_{\infty}$, and the volume filling 
                factor $F_{cl}$. For each model, we also calculate synthetic spectra in the UV 
                (900-2000 \AA), optical (3500-7000 \AA), and near-IR (10000-40000 \AA) regions. To 
                facilitate comparison with observations, the synthetic spectra can be rotationally 
                broadened using ROTIN3, by covering \vsini\ velocities between 10 and 350 km s$^{-1}$ 
                with steps of 10 km s$^{-1}$.}
              { We also present the results of the reanalysis of $\epsilon$~Ori using our grid 
                to demonstrate the benefits of databases of precalculated models. Our analysis 
                succeeded in reproducing the best-fit parameter ranges of the original study, although our results favor the higher end of the mass-loss range and 
                a lower level of clumping. Our results indirectly suggest that the resonance lines 
                in the UV range are strongly affected by the velocity-space porosity, as has been suggested 
                by recent theoretical calculations and numerical simulations.}


\titlerunning{Large grids of model atmospheres}
\authorrunning{Zsarg\'o et al.}

\maketitle

\section{Introduction}
\label{sec:intro}

The self-consistent 
analysis of spectral regions from the X-ray to the IR is now possible because of the fertile combination of the large amount of observational data and the availability 
of sophisticated stellar atmosphere codes, such as CMFGEN \citep{hil13,hil98,hil99,pue16}, PHOENIX 
\citep{hau92}, TLUSTY \citep{hub95}, WM-BASIC \citep{pau01}, FASTWIND \citep{san97,pul05,riv11,
car16}, and the Potsdam Wolf-Rayet (W-R) code \citep[PoWR,][]{gra02,ham04,san15}. As a result, significant 
advances have been made in our understanding of the physical conditions in the atmospheres and 
winds of massive stars. In parallel with these advances, stellar atmosphere codes were also 
improved and became even more sophisticated.

For example, early far-UV observations \citep{ful00} showed inconsistencies between 
the optical effective temperature scale and the temperature that was implied by the observed wind 
ionization. Studies by \cite{mar02} and others showed that the neglect of line blanketing in the 
atmospheric models resulted in a systematic overestimate of the effective temperature derived from 
optical H and He lines. After these effects were introduced into the stellar atmosphere models, the 
inconsistencies were eliminated. Studies by \cite{cro02}, \cite{hil03}, and \cite{bou03}, who 
simultaneously analyzed $FUSE$, $HST$, and optical spectra of O stars with the corrected models, 
were able to derive consistent effective temperatures using a wide variety of diagnostics. 

Another crucial development was the recognition of the important effects of the line-deshadowing 
instabilities (LDI or clumping) on the spectral analyses of O, B, and W-R stars. For example, \cite{cro02} 
and \cite{hil03} were unable to reproduce the observed \ion{P}{V} $\lambda\lambda$1118-1128 profiles 
when they used mass-loss rates derived from the analysis of H$\alpha$. The only way in which the \ion{P}{V} and 
H$\alpha$ profile discrepancies could be resolved were either to assume substantial clumping or to 
use unrealistically low phosphorus abundances. As a consequence of the introduction of clumping, 
the mass-loss rates were lowered by significant factors (from 3 to 10). The analysis of the  high-resolution X-ray 
observations of the $Chandra$ and $XMM-Newton$ X-ray observatories provided further support to the 
idea of a reduced mass-loss rate. The strong emission lines produced by H and He-like ions of O, Ne, 
Mg, and Si are far less strongly absorbed by the cold wind than previously expected 
\citep[e.g.,][]{cas01, kah01}. For example, \citet{coh05, coh10}, \citet{coh14}, and \cite{gag05} analyzed the 
bound-free absorption by the wind in massive stars observed by $Chandra$ and found mass-loss rates 
that were up to ten times lower than previous estimates. 

Including X-ray regions into self-consistent stellar atmospheric models also resulted in the 
introduction of the non-vacuum interclump medium in the wind models. The only way \cite{zsa08} 
were able to reproduce the strong \ion{O}{VI} and \ion{N}{V} lines in the far-UV (FUV) spectra of $\zeta$ Pup 
were to abandon the assumption of vacuum between clumps. Now, we know that these super-ions are 
primarily produced by Auger ionization by X-rays in the rarified interclump medium 
\citep[see also][]{pue16}. Furthermore, ideas to explain the low wind absorption 
on the X-ray emission lines by \cite{osk04} that invoked optically thick clumping, although they 
were refuted by \citet{owo06} and \citet{sun12}, led to the recognition of the importance of velocity-space 
porosity (vorosity) effects on strong UV lines \citep[][and references therein]{sun14, sun18}. 
 
Unfortunately, the ever improving sophistication of the model calculations also means that 
running these codes and performing a reliable analysis is rather difficult and requires much 
experience, which many researchers do not have enough time to gain. It is therefore
useful to develop databases of precalculated models. Such databases will free up valuable time 
for astronomers who could study stellar atmospheres with reasonable accuracy but with fewer time-consuming simulations. Furthermore, these databases will not only accelerate the 
studies of the large number of observed spectra that are in line for analysis, but also ensure that 
they are made in a uniform manner (e.g., using the same atomic data). This uniformity will also 
help to identify and correct possible shortcomings in the codes that were used to produce the database. 

The basic parameters of such databases of precalculated models are the surface temperature (\Teff), 
the stellar mass ($M$), and the surface chemical composition. An adequate analysis of massive stars 
also has to take into account the parameters associated with the stellar wind, such as the terminal 
velocity ($V_\infty$), the mass-loss rate ($\Mdot$), and the clumping. When the 
variations of all necessary parameters are taken into account, the number of precalculated models that are needed will increase exponentially. Production of such databases is therefore only possible using clusters 
of computers or supercomputers.

A few databases of synthetic stellar spectra are currently available, but only with a few dozen 
or some hundred stellar models \citep[see, e.g.,][]{fie15, ham04, pal10, hai19}. 
On the other hand, we here generate a database with tens of thousands of models \citep{zsa17} 
that will be publicly available in about January 2021. It will be impossible to manually 
compare an observed spectrum with this many model calculations. It is therefore imperative 
to develop tools that allow the automation of this process, but without compromising the quality of 
the fitting. \cite{fie18} presented FIT\textit{spec}, the  first of these tools, that searches our 
database for models that best fit the observed spectrum in the optical. It uses the Balmer lines 
to measure the surface gravity (\logg) and the equivalent width ratios of the \ion{He}{II} and 
\ion{He}{I} lines to estimate the surface temperature (\Teff). 

In this article we describe the development state of our grid of precalculated models and the results of a 
test analysis to verify the usefulness of the grid. In \S~\ref{sec:cmfgen} and \S~\ref{sec:cmf_flux} 
we briefly describe the stellar atmosphere code (CMFGEN) that we use to produce our models. Next,
in \S~\ref{sec:models} we describe our model grid, and \S~\ref{sec:simtest} presents a simple 
test analysis to demonstrate the benefits that our grid offers. Finally, in \S~\ref{sec:disc} and 
\ref{sec:concl} we summarize our conclusions.

\section{CMFGEN}
\label{sec:cmfgen}
CMFGEN \citep{hil13, hil98, hil99, pue16} is a sophisticated and widely-used nonlocal thermal equilibrium (non-LTE) stellar 
atmosphere code. It models the spectrum from the FUV to the radio wavelength range and has been used 
successfully to model O and B stars, W-R stars, luminous blue variables, and even supernovae. 
Recently, an experimental version of the code has been developed that included the X-ray region 
in the analysis \citep{zsa08, pue16}, but this version has not yet been made public. CMFGEN
determines the temperature, the ionization structure, and the level populations for all elements 
in the stellar atmosphere and wind. It solves the spherical radiative transfer equation in the 
comoving frame in conjunction with the statistical equilibrium and radiative equilibrium
equations. The hydrostatic structure can be computed below the sonic point, allowing the
simultaneous treatment of spectral lines formed in the atmosphere, in the stellar wind, and in the 
transition region between the two. Such features make it particularly well suited to the study of 
massive OB stars with winds. However, there is a price for this sophistication: a typical
CMFGEN simulation takes anywhere between 24 and 36 hours of microprocessor time to complete.

For atomic models, CMFGEN uses the concept of super levels, by which levels of similar 
energies are grouped together and treated as a single level in the statistical equilibrium equations
\citep[see][and references therein for more details]{hil98}. The atomic model used in this project includes 
37 explicit ions of the different elements, which are summarized in Table~\ref{tab:superlevels} 
together with the levels and super levels included in our model. The atomic data references are 
given in \cite{her04}.
%
\begin{table*}
\begin{center}
\caption{Super levels and levels for the different ionization stages included in the models.
\label{tab:superlevels}}
\begin{tabular}{lllllllll}
\multicolumn{1}{c}{Element} & \multicolumn{1}{c}{I} &
\multicolumn{1}{c}{II} & \multicolumn{1}{c}{III} & \multicolumn{1}{c}{IV} & 
\multicolumn{1}{c}{V} & \multicolumn{1}{c}{VI} & \multicolumn{1}{c}{VII} &
\multicolumn{1}{c}{VIII}\\
\hline
H & 20/30 & 1/1 & \nodata & \nodata & \nodata & \nodata & \nodata & \nodata\\
He & 45/69 & 22/30 & 1/1 & \nodata & \nodata & \nodata & \nodata & \nodata\\
C & \nodata & 40/92 & 51/84 & 59/64 & 1/1 & \nodata & \nodata & \nodata\\
N & \nodata & 45/85 & 41/82 & 44/76 & 41/49 & 1/1 & \nodata & \nodata\\
O & \nodata & 54/123 & 88/170 & 38/78 & 32/56& 25/31 & 1/1 &\nodata\\
Si & \nodata & \nodata & 33/33 & 22/33 & 1/1 & \nodata & \nodata & \nodata\\
P & \nodata & \nodata & \nodata & 30/90 & 16/62 & 1/1 & \nodata &\nodata\\
S & \nodata & \nodata & 24/44 & 51/142 & 31/98 & 28/58 & 1/1 & \nodata\\
Fe & \nodata & \nodata & 104/1433 & 74/540 & 50/220 & 44/433 & 29/153 & 1/1\\ 
\hline
\end{tabular}
\end{center}
\end{table*}
 
\begin{figure}
\centering
\includegraphics[width=0.95\linewidth]{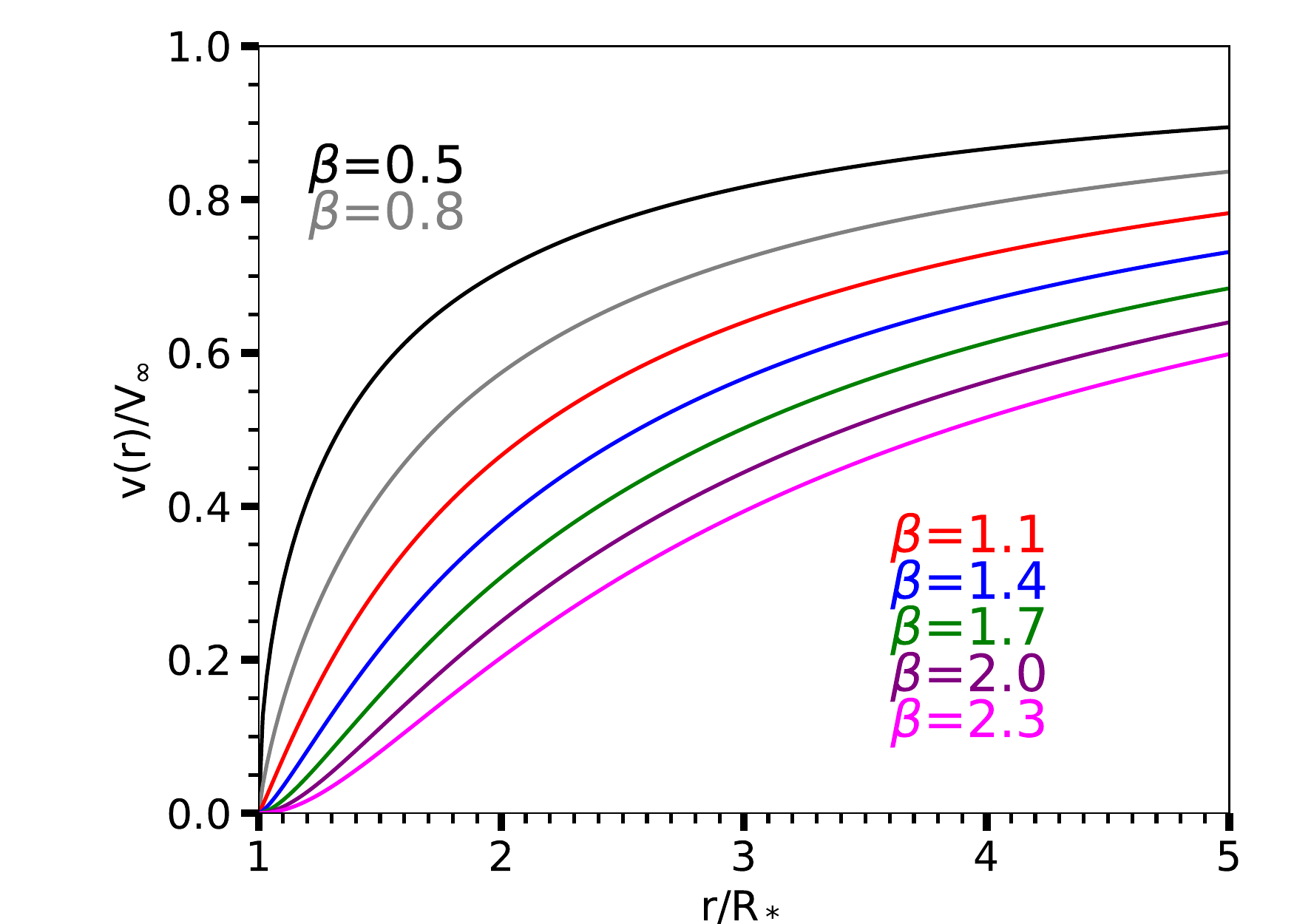}
\caption{Examples of $\beta$-type velocity laws. The curves and the corresponding $\beta$ values are color-coded.}
\label{f:beta_law}
\end{figure}

To model the stellar wind, CMFGEN requires values for the mass-loss rate \Mdot, terminal 
velocity \Vinf, $\beta$ parameter, and the volume filling factor \Fcl.  The profile of the 
wind speed is modeled by a $\beta$-type law \citep{CAK},
\begin{equation}
\label{e:beta_law}
v(r)=V_{\rm\infty}\left(1-\frac{r}{R_{\star}}\right)^{\beta}.
\end{equation}
The $\beta$ parameter controls how fast the stellar wind is accelerated to reach the terminal 
velocity (see Fig.~\ref{f:beta_law}), while the volume filling factor \Fcl\ is the standard 
method with which the atmospheric models introduce optically thin clumping in the wind 
\citep[see, e.g.,][and references therein]{sun14}. In short, \Fcl\ gives the average fraction of 
a volume element that is filled with material at the outer regions of the wind, where $v(r)\sim$\Vinf. 
Between these regions filled with material, called the clumps, CMFGEN assumes a vacuum. Because of this assumption, we should not expect that our models can reproduce the strong UV lines of 
the super-ions \ion{O}{VI} and \ion{N}{V} (see \S\ref{sec:intro} for an explanation). Because the 
continuity equation requires that the average wind density is defined as 
$\langle\rho\rangle =$\Mdot$/[4\pi r^2 v(r)]=\rho _{\rm cl}f_{\rm cl}(r)$, the volume filling 
factor also tells us how the degree of overdensity in the clumps compares to 
$\langle\rho\rangle$. In CMFGEN the radial distribution of the volume filling factor is described 
by the relation
\begin{equation}
\label{e:clumping}
f_{\rm cl}(r)=F_{\rm cl}+(1-F_{\rm cl})\cdot\exp\left[-\frac{v(r)}{V_{\rm cl}}\right],
\end{equation}
where the parameter \Vcl\ controls how fast \Fcl\ is reached in the wind. The lower \Vcl\ , the 
faster and closer to the stellar surface does the volume filling factor reach the value of \Fcl. 
During the generation of our models in the grid, we used the generic value \Vcl$=0.1$\Vinf.
%
%

\subsection{Synthetic spectra}
\label{sec:cmf_flux}
The auxiliary program CMF\_FLUX of the CMFGEN package \citep{hil13} computes the synthetic observed
spectrum in the observer's frame, which is one of the most important outputs of our models. To 
simulate the effects of rotation on the spectral lines, the synthetic spectra can also be rotationally 
broadened using the program ROTIN3, which is part of the TLUSTY package \citep{hub95}.

For each model in the grid, we calculate the normalized spectra in the UV (900-3500\AA), optical 
(3500-7000 \AA), and IR (7000-40 000 \AA) range. Then, we can apply rotation by sampling the range 
between 10 and 350 km s$^{-1}$ with steps of 10 km s$^{-1}$. 

\section{Model grid}
\label{sec:models}

In order to properly constrain the input parameters, we used the evolutionary tracks and isochrones of \cite{eks12} calculated with solar metallicity ($Z=0.014$) at the zero-age main sequence 
(ZAMS). They computed 24 different evolutionary tracks with initial masses between 0.8 and 120~\Msun\ as well as 37 isochrones for both rotating and stationary models. The rotating models start on the ZAMS with an equatorial rotational velocity V$_{\rm eq}$=0.4V$_{\rm crit}$ that is later evolved as the star loses angular momentum through mass loss or magnetic braking. The differential rotation within these stars gives rise to meridional circulation that enhances the surface chemical composition by moving CNO-processed material from the interior to the surface. The theory of rotation is described in a series of papers by the Geneva group  \cite[][and references therein]{mae99, mae00}. We refer to these papers for more information. 

The predictions of \cite{vin01} were used to calculate mass loss for normal early-type stars and lower mass W-R stars, and the recipe of \cite{gra08} was used for massive W-Rs. The evolution was computed until the end of the central carbon-burning phase, the early AGB phase, or the core helium-flash for massive, intermediate, and low- and very low-mass stars, respectively. The initial abundances were those deduced by \cite{asp09}, which best fit the observed abundances of massive stars in the solar neighborhood. For each track, \cite{eks12} calculated 400 evolutionary stages and provided the luminosity $L$, the surface temperature \Teff, and the mass $M$ together with mass-loss rates \Mdot, the surface chemical composition, and the parameters that describe 
the stellar interior \citep[see][for more details]{eks12}. This covers all parameters that our 
models need, except for those that describe the structure of the wind. For the stellar radius and 
\logg\ we calculated values that are consistent with their published luminosities $L$, surface
temperatures \Teff, and masses $M$ by using the Stefan-Boltzmann law and the assumption of 
spherical symmetry.

The elements included in our models are H, He, C, N, O, Si, P, S, and Fe. The abundances of H, He, 
C, N, and O were taken from the tables of \cite{eks12}. For consistency, we assumed solar metallicity 
as reported by \cite{asp09} for Si, P, S, and Fe in all models. 

\begin{figure*}
\includegraphics[width=0.52\textwidth]{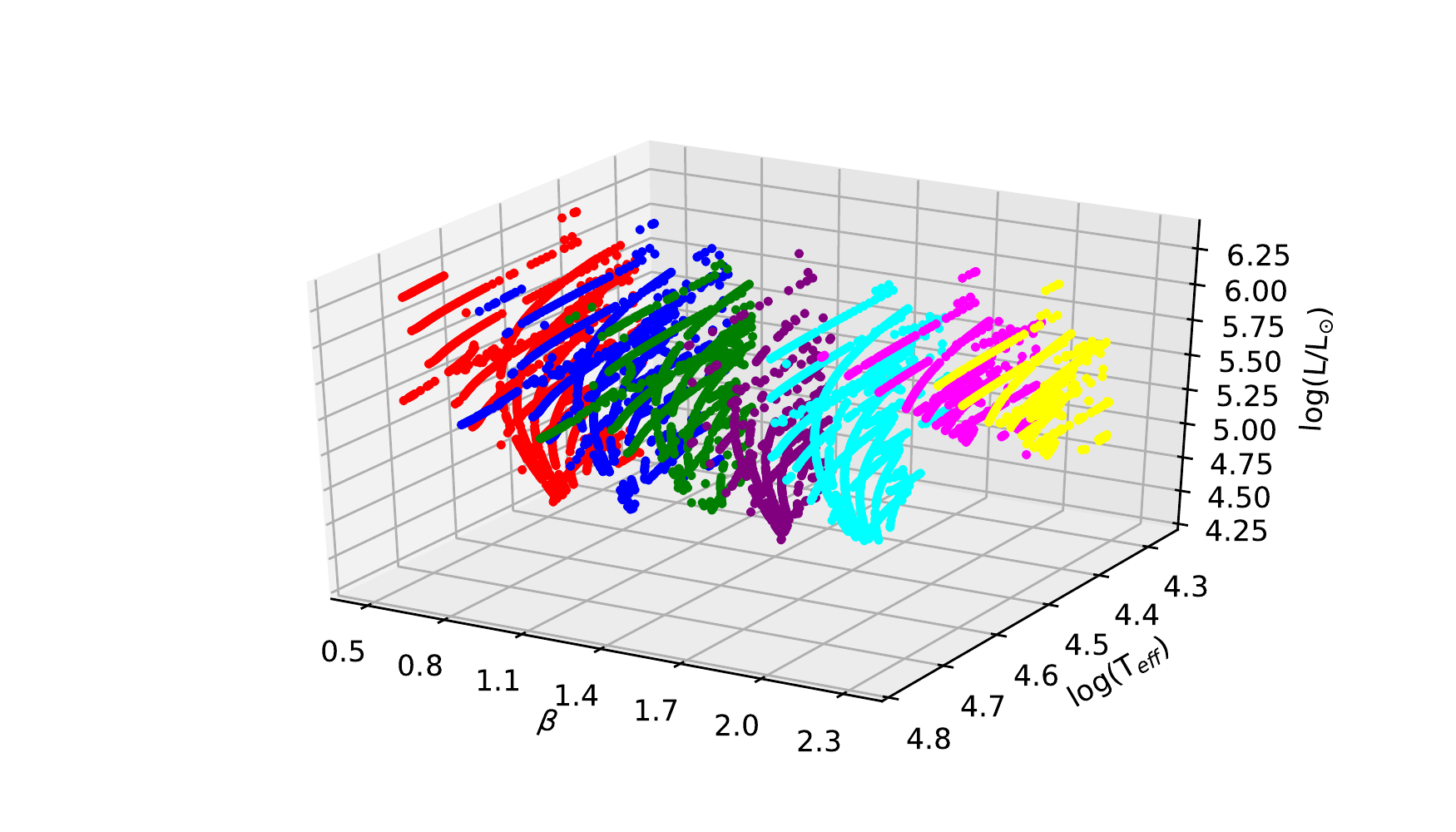}
\includegraphics[width=0.52\textwidth]{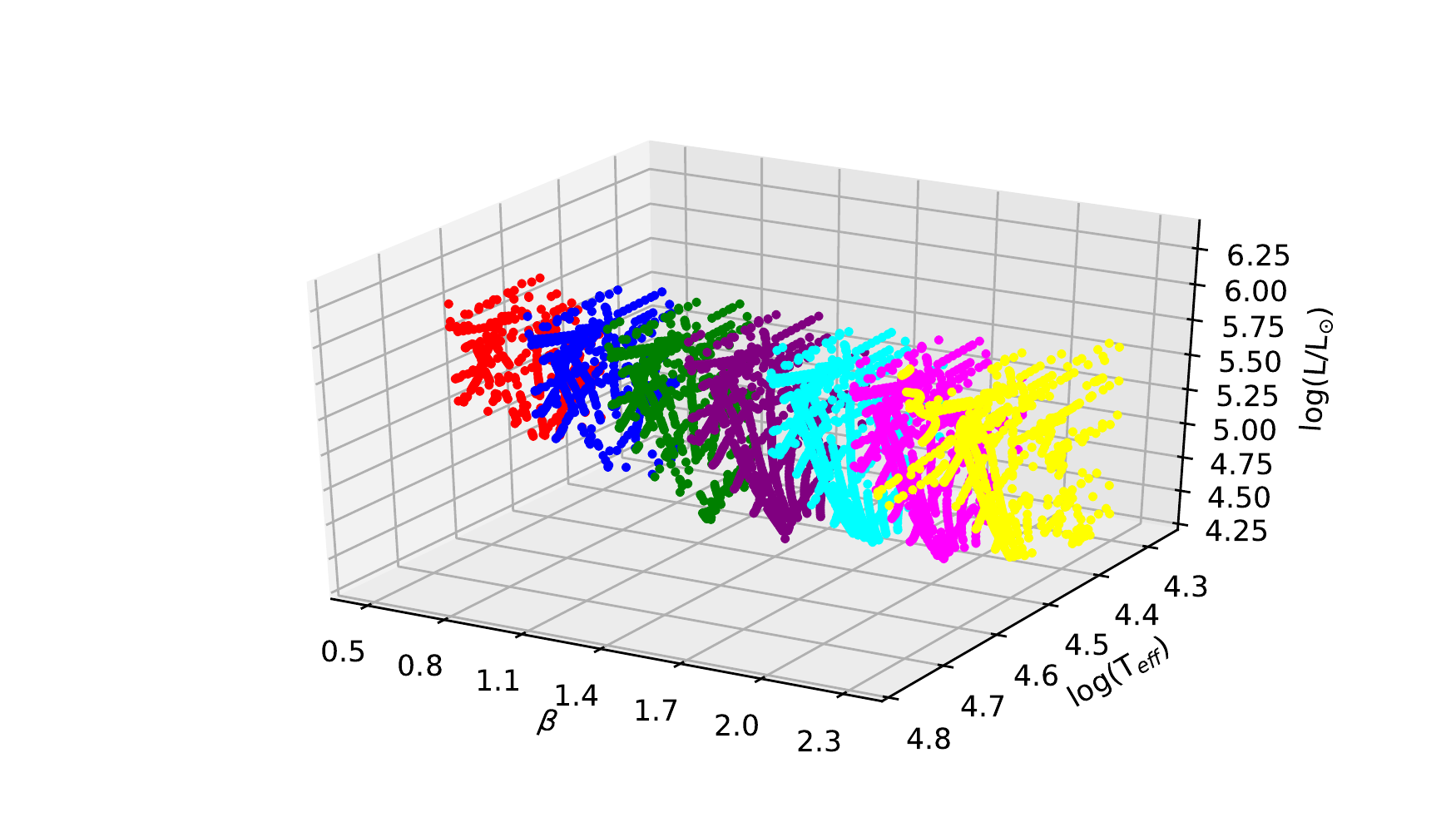}
\includegraphics[width=0.52\textwidth]{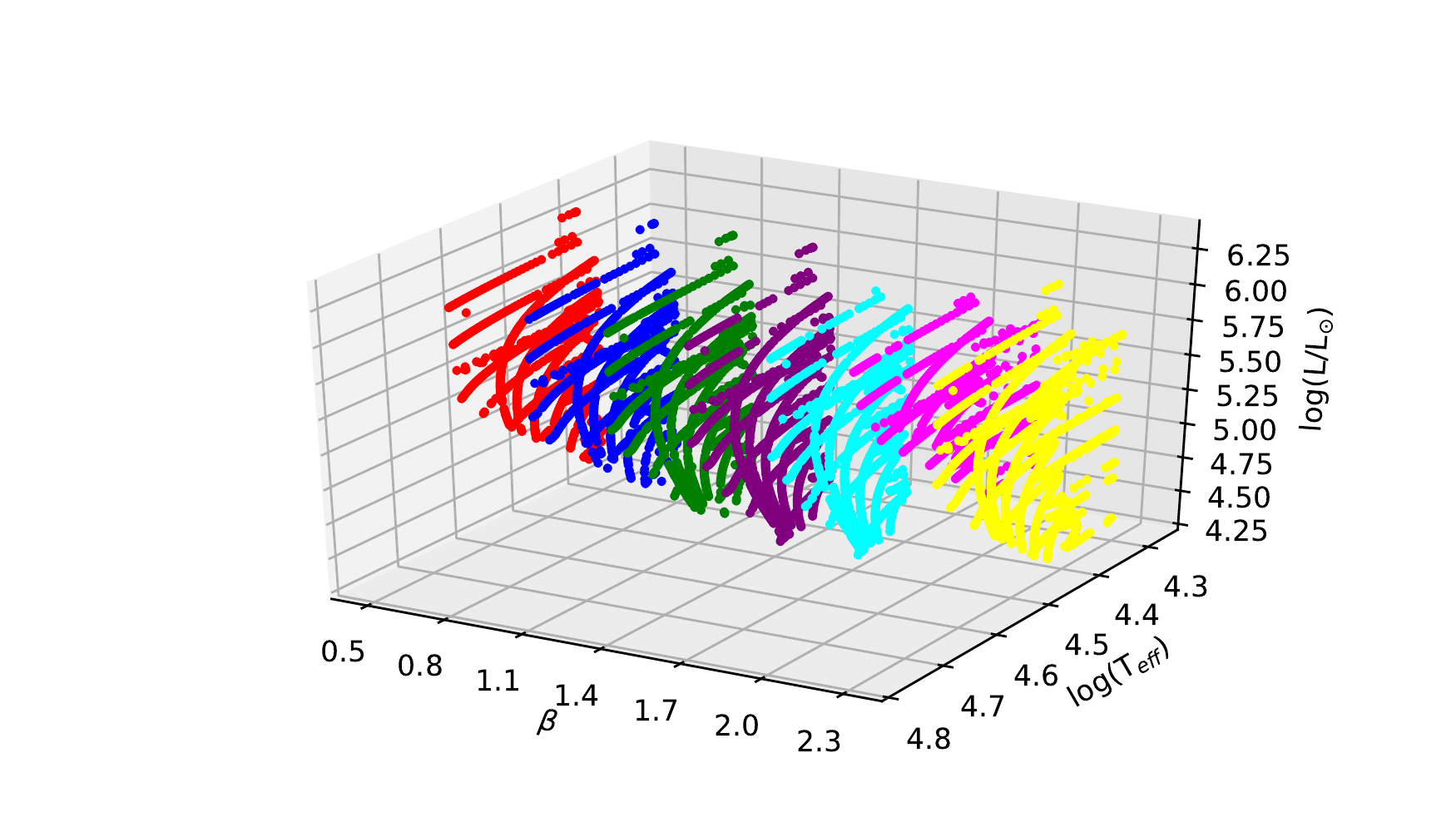}
\includegraphics[width=0.52\textwidth]{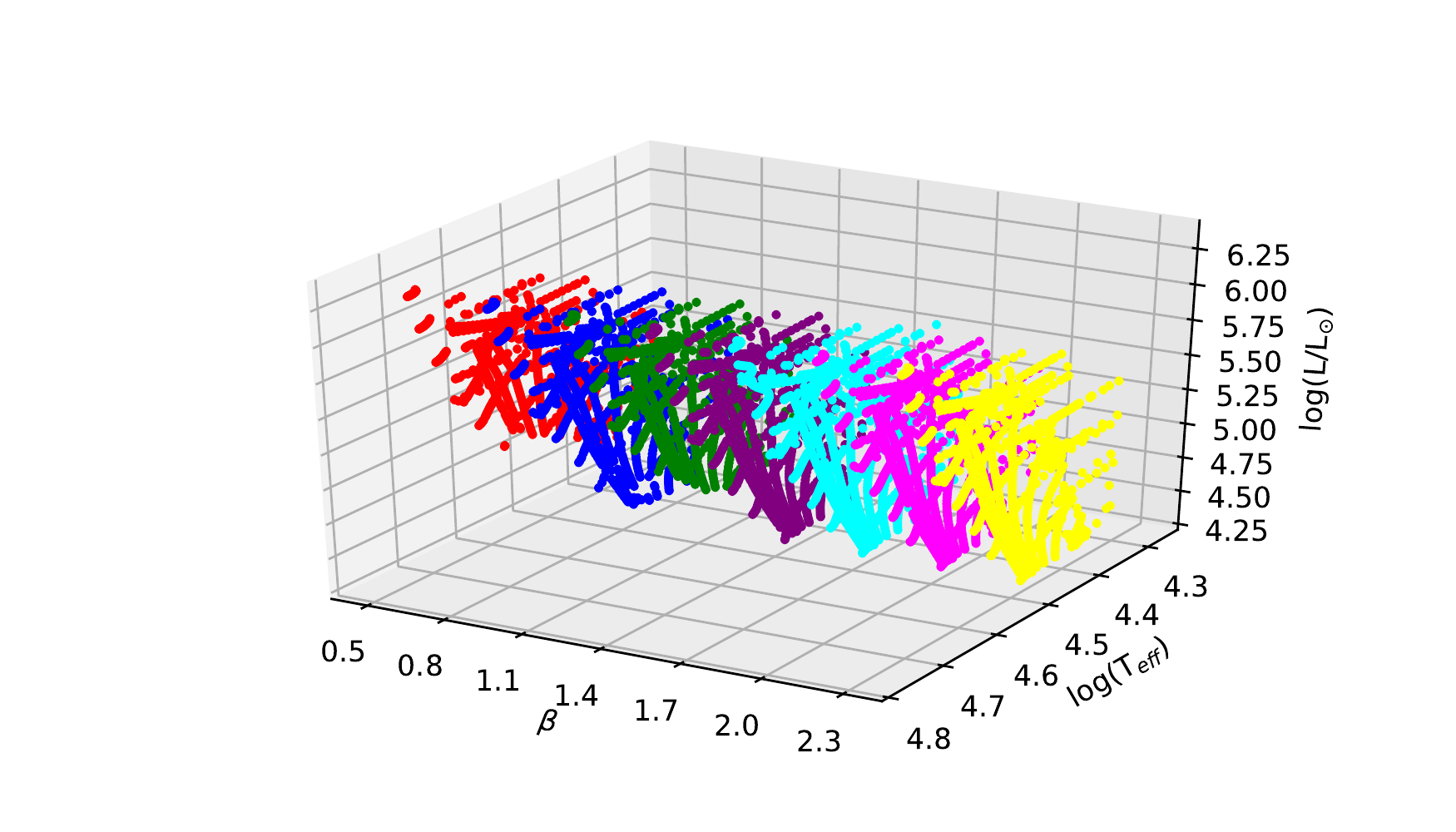}
\includegraphics[width=0.52\textwidth]{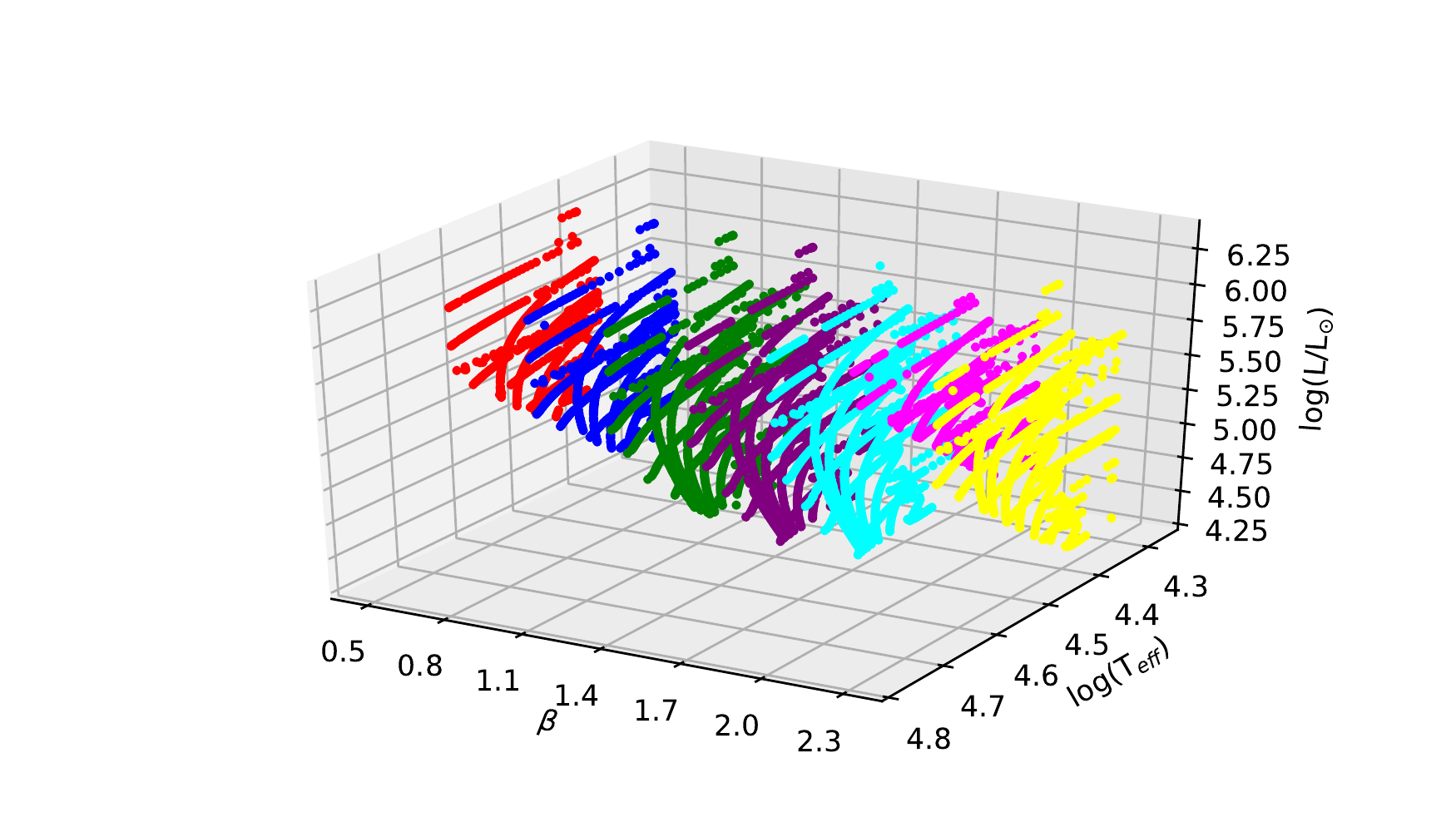}
\includegraphics[width=0.52\textwidth]{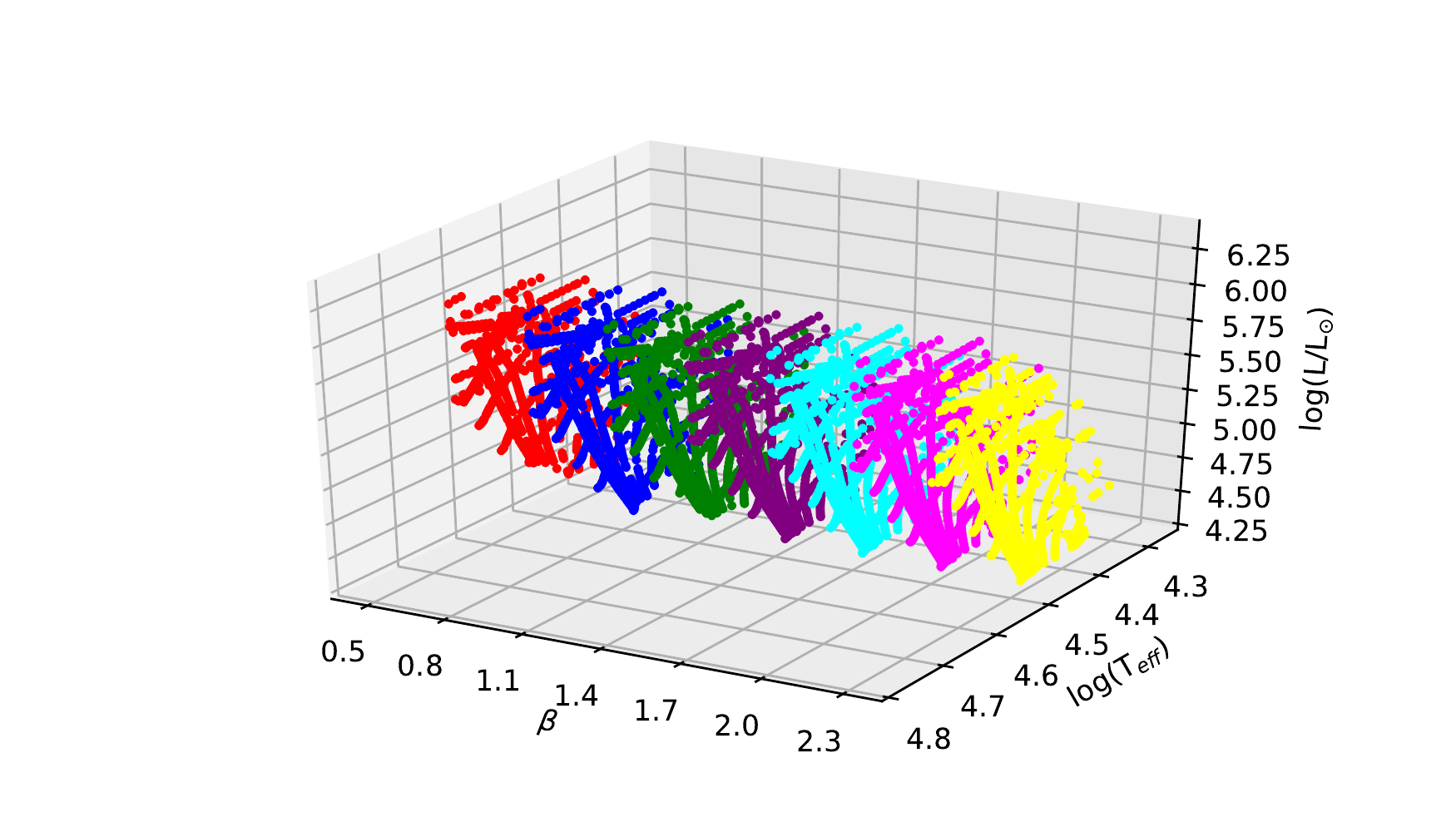}
\caption{Hyper-cube plane formed by 3D datacubes. The dimensions of this plane are the different values of the volume filling factor (\Fcl= 0.05, 0.3, and 0.6 from top to bottom) and the two metallicities (right column: solar enhanced by rotation; left column: solar without rotational enhancement). Each datacube contains seven H-R diagrams formed by the models in our grid. We used color-coding to help visualize the H-R diagram corresponding to a $\beta$ parameter.}
\label{f:cubos}
\end{figure*}

\begin{figure*}
\centering
\includegraphics[width=0.8\textwidth]{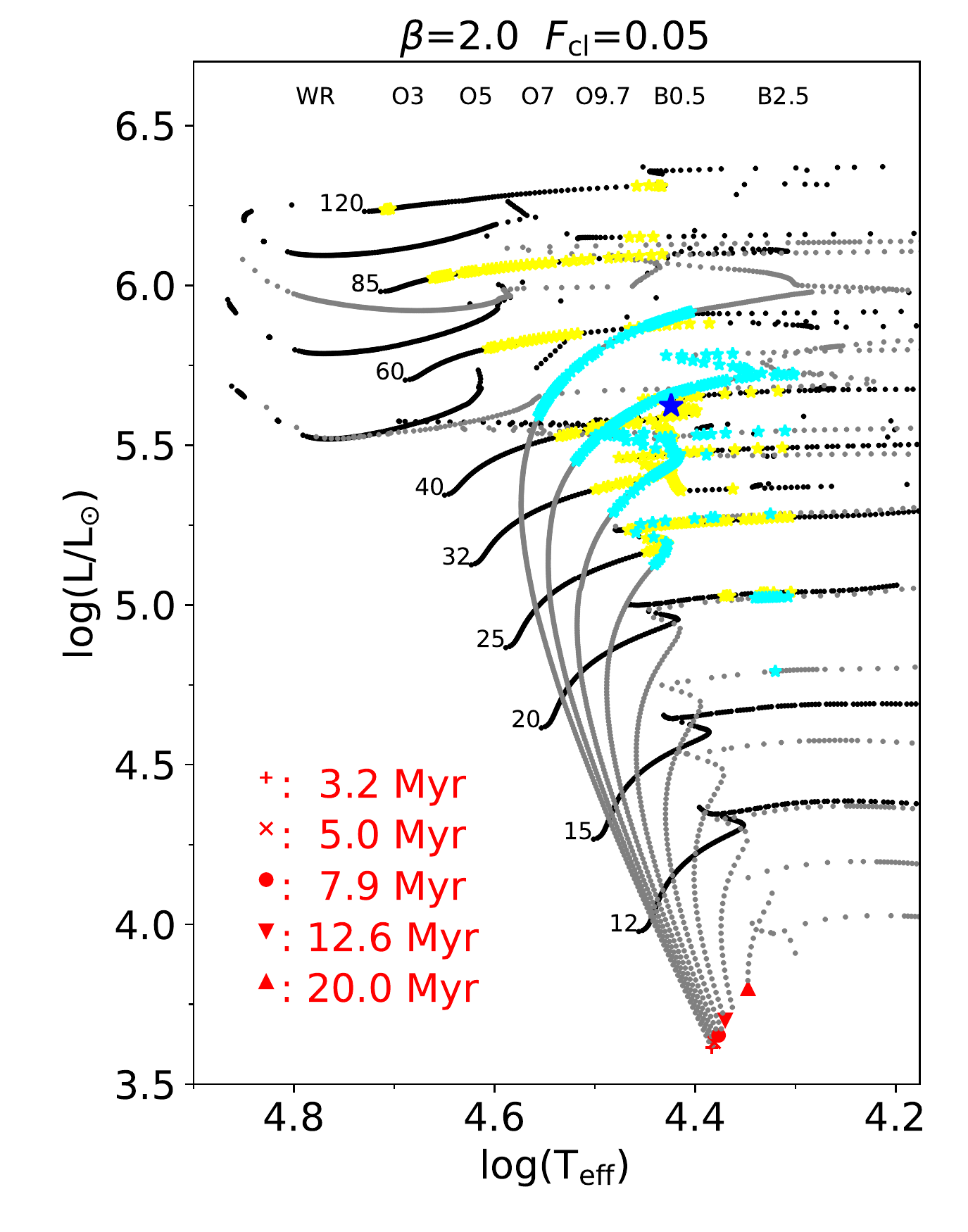}
\caption{Hertzsprung-Russel diagram for $\beta$=2, \Fcl=0.05, and solar composition without rotational enhancement.  The black and gray symbols mark the evolutionary tracks and isochrones, respectively, calclulated by \cite{eks12}. We use the stellar parameters from these models, together with our choice of wind parameters, to create the CMFGEN atmospheric models in our grid. The yellow and cyan stars indicate evolutionary track and isochrone models, respectively, for which we have already generated  an atmospheric model. The large blue star shows the location of our best-fit model to the observation of $\epsilon$ Ori. The numbers left of the evolutionary tracks are the initial masses of the tracks (in  \Msun). They also mark the approximate location of the ZAMS. The red symbols below selected isochrones indicate their ages  (in  10$^6$ years). At the top of the figure we indicate the approximate locations of selected spectral classes. \label{f:HRDs_0.05nrot}}
\end{figure*}

\begin{figure*}
\centering
\includegraphics[width=0.8\textwidth]{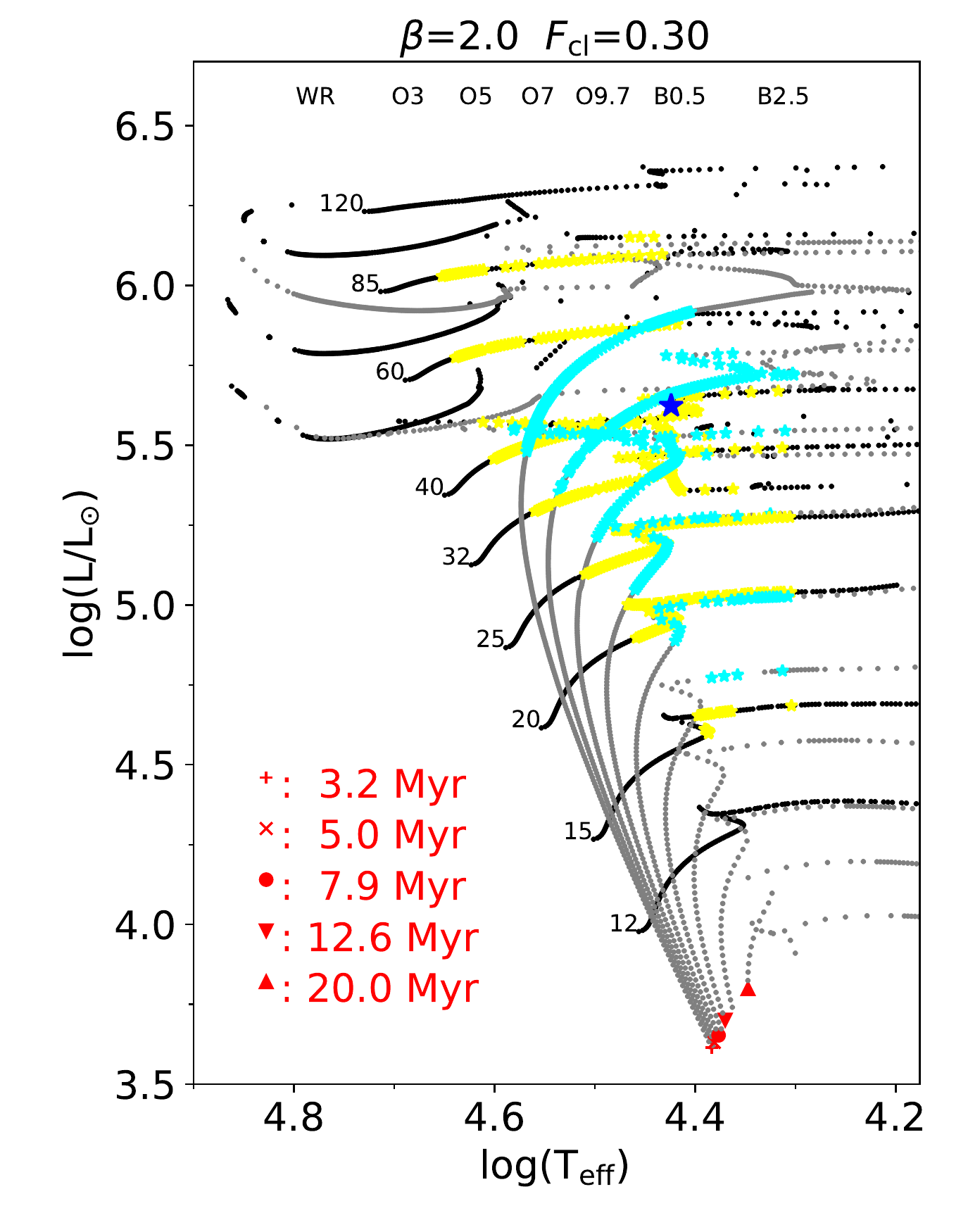}
\caption{Same as Fig.~\ref{f:HRDs_0.05nrot}, but for models with \Fcl$=0.3$ and solar chemical 
composition without rotational enhancement.}
\label{f:HRDs_0.3nrot}
\end{figure*}

\begin{figure*}
\centering
\includegraphics[width=0.8\textwidth]{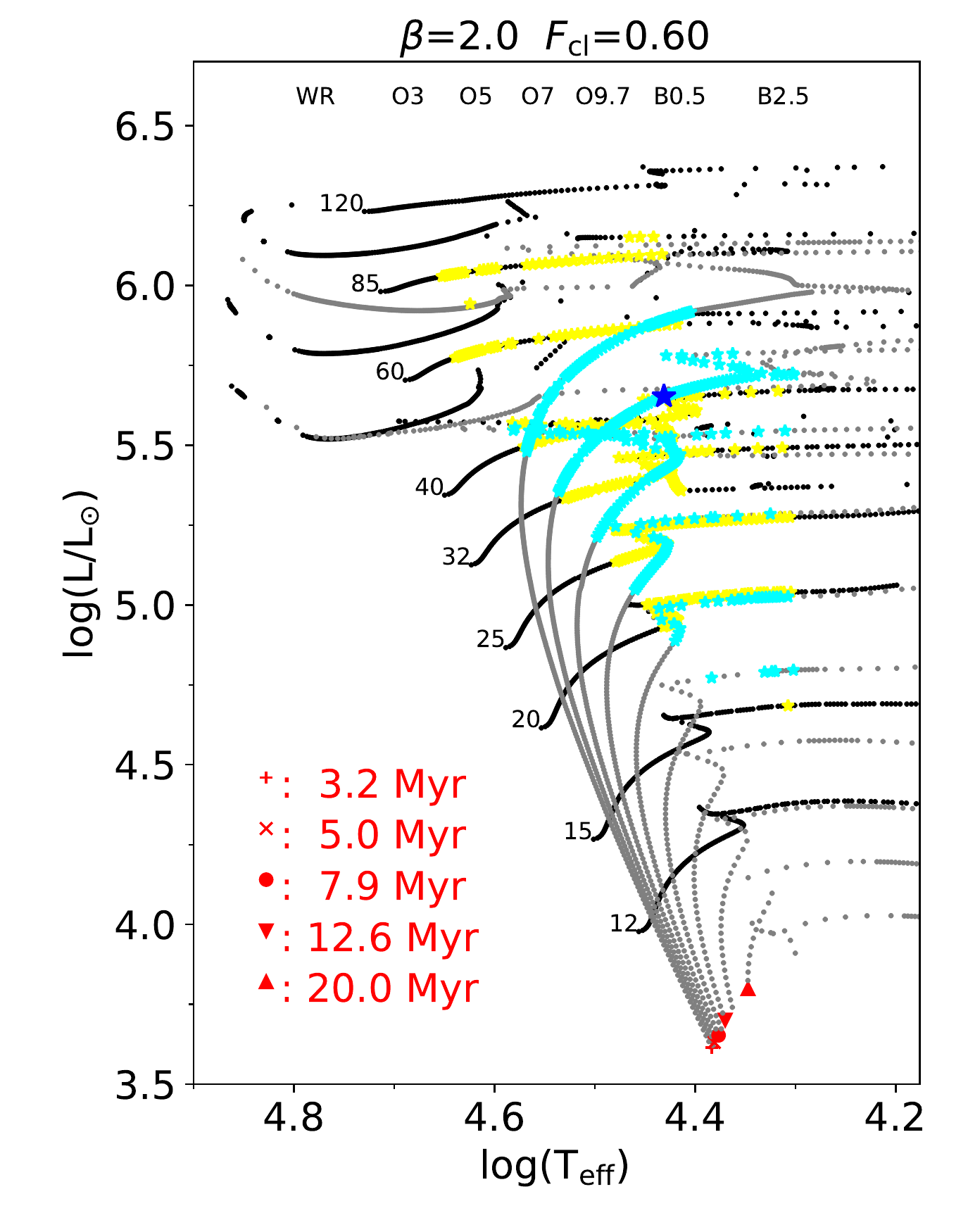}
\caption{Same as Fig.~\ref{f:HRDs_0.05nrot}, but for models with \Fcl$=0.6$ and solar chemical 
composition without rotational enhancement.}
\label{f:HRDs_0.6nrot}
\end{figure*}

We only had to improvise to describe the wind structure, that is, to
determine values for the terminal velocity \Vinf, $\beta$ parameter, and volume filling factor \Fcl. These parameters have no relevance in the evolutionary calculations, therefore no 
values were reported in \cite{eks12}, but they are very important to reproduce observed spectra. 
The terminal velocity is thought to be related to the escape velocity for hydrodynamical 
considerations. We therefore used $V_{\rm esc}$ to estimate its values (where $V_{\rm esc}$ has 
its usual meaning). For $\beta$ and \Fcl, we simply used the range of values reported in previous 
studies of massive stars to define the grid.

The grid is organized as hyper-cube data of dimensions that correspond to \Vinf, \Fcl, and the 
metallicity, as illustrated in Fig.~\ref{f:cubos}. For \Vinf\ we plan to use two values, a low- 
(\Vinf$=1.3V_{\rm esc}$) and a high- (\Vinf$=2.1V_{\rm esc}$) velocity model, to cover the range 
of terminal velocities reported in the literature. However, up to the 
preparation of this paper we have only generated models for the higher velocity. The hyper-cube 
in Fig.~\ref{f:cubos} therefore has only two dimensions, given by \Fcl\ and the metallicity. Each value of 
the volume filling factor (\Fcl=0.05, 0.3, 0.6, 1.0) combined with one of the two possible values 
of the metallicity (solar without rotational enhancement and solar enhanced by rotation) generates 
a datacube that contains a 3D space of models with dimensions that are defined by $\beta$, 
\Teff, and $L,$ as illustrated by Fig.~\ref{f:cubos}. The planes generated by each $\beta$ parameter 
(with $\beta$=0.5, 0.8, 1.1, 1.4, 1.7, 2.0, 2.3) within a datacube are the H-R diagrams 
(see, e.g., Figs.~\ref{f:HRDs_0.05nrot}--\ref{f:HRDs_0.6nrot}), where the values of $L$ and 
\Teff\ are restricted by the evolutionary tracks.  

In Figs.~\ref{f:HRDs_0.05nrot} to \ref{f:HRDs_0.6nrot} (see also Fig.~\ref{f:HRDs_appx} in the 
appendix) we show a sample of the H-R diagrams formed by our models in the datacubes. Obviously, during the generation of our grid we only consider the region in the complete H-R diagram that is relevant for massive stars with line-driven winds (O/B-types and W-Rs), therefore only the upper left part of the full H-R diagram is covered by our models. We do not generate CMFGEN model for each evolutionary stage that is shown in the figures (we do not have the computing resources for this), only for the models that have \Mdot>10$^{-8}$\Msun yr$^{-1}$. The large blue stars in the figures mark the location of the models that best-fit the observations of $\epsilon$ Ori. We discuss the implication of the position of these models in \S\ref{sec:disc}.  

The coverage of the relevant part of the H-R diagram for the wind parameters is quite diverse, and 
generally better for intermediate masses than for the extremes. It is also quite clumpy, with clusters of models separated by gaps; this is not ideal for spectral analysis. The reason for this is technical, and it is the consequence of our strategy of optimizing the production of our models. CMFGEN, like  any other code that relies on iterative methods to solve a nonlinear system of equations, needs good initial estimates to achieve secure convergence in reasonable time. Because it is difficult to produce such estimates, the first models (the seed models) in a new region of the parameter space can be quite troublesome. They often fail to converge and can run 3-4 times longer than a normal model. It is therefore wise to minimize the number of seed models during the generation of the models and grow the grid around existing seed models that can provide good initial conditions for the next generations of models. We use this strategy to optimize the usage of the computing facilities; however, this results in the clumpy distribution of models described above.

\begin{figure*}
\centering
\includegraphics[width=0.8\textwidth]{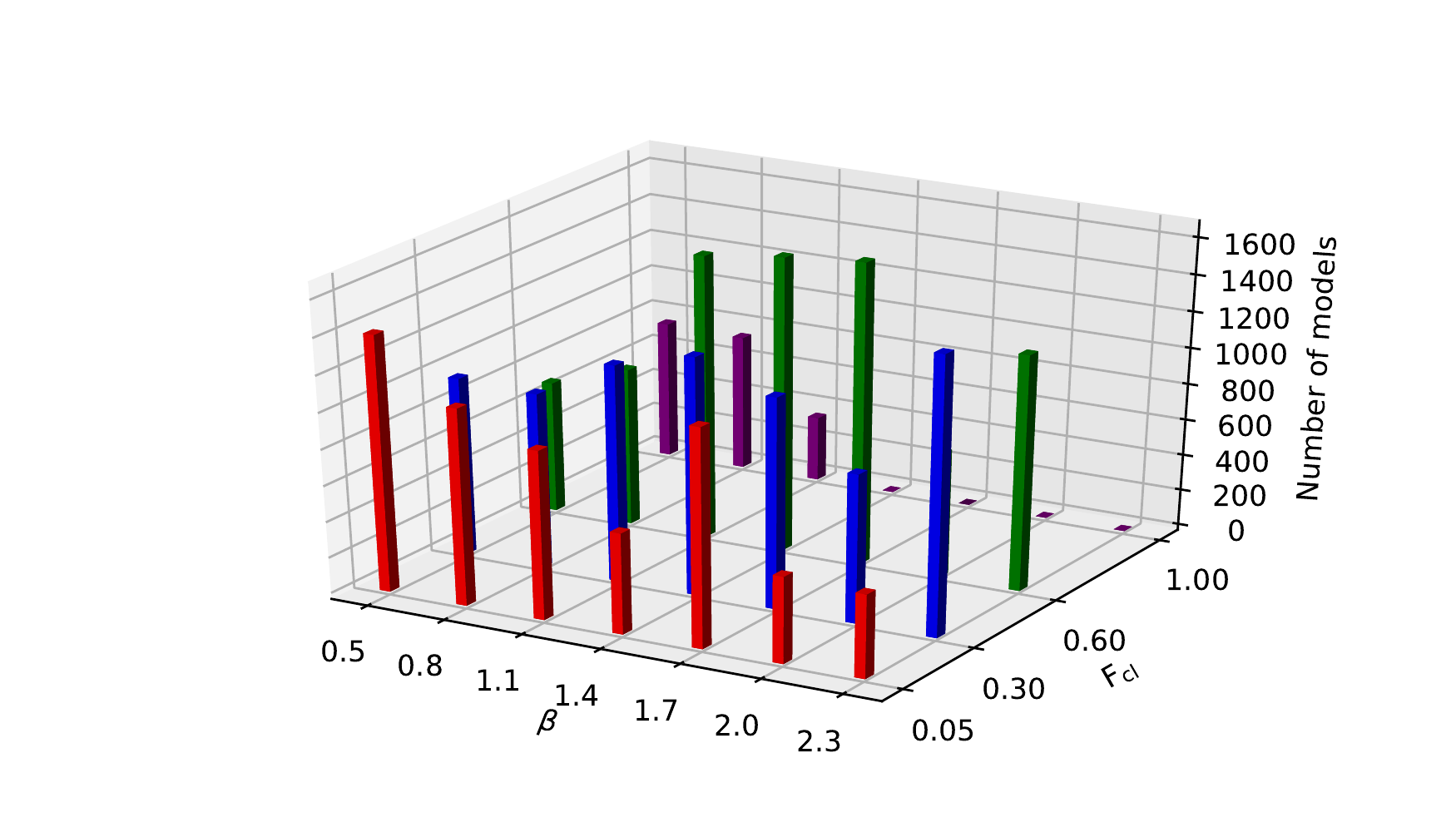}
\caption{Distribution of the 24,423 models with no rotational enhancement that are currently available as functions of wind parameters. The color-coding is to help visualize the models that correspond to the same \Fcl\ volume filling factor.}
\label{f:dist_nrot}
\end{figure*}

\begin{figure*}
\centering
\includegraphics[width=0.8\textwidth]{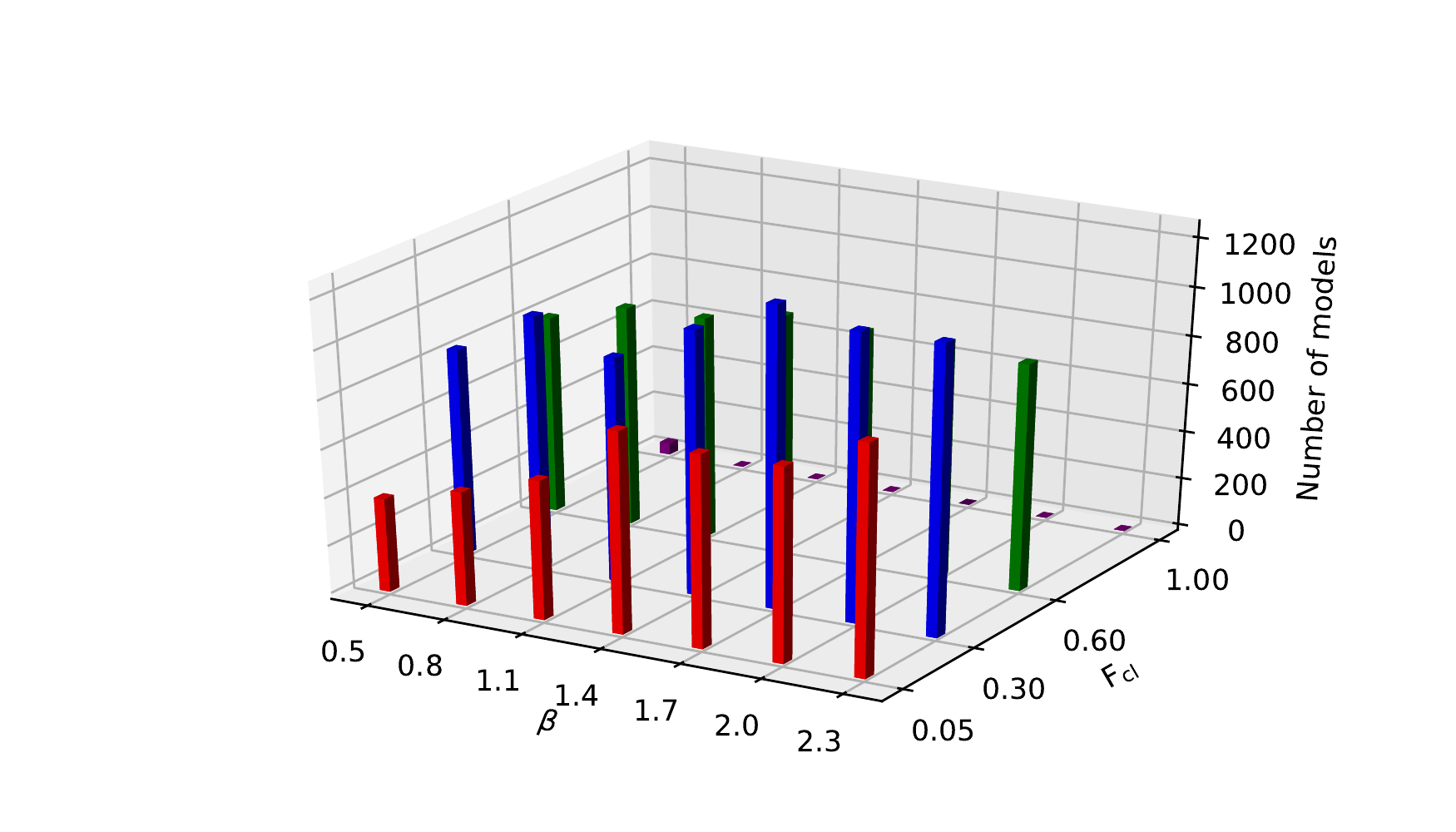}
\caption{Same as Fig.~\ref{f:dist_nrot}, but for the 18,917 models with rotational enhancement that are currently available in the grid.}
\label{f:dist_rot}
\end{figure*}

We do not show any models for \Fcl=1. The reason for this is clear from Figs.~\ref{f:dist_nrot} and \ref{f:dist_rot}, which show the number distribution of the models as functions of the \Fcl\ and $\beta$ parameters. Our coverage for all other \Fcl\ volume filling factors is decent and we have many models (43,340), but not for \Fcl=1. The coverage for this volume filling factor is so low that it is pointless to show any H-R diagrams.  Fortunately, it is highly unlikely that modelers encounter a wind that has no density structures (clumps) at all. Calculations with \Fcl=1 therefore had low priorities so far, hence the low number of models. Nevertheless, these models will be produced and will be available in the completed grid.
 
Despite the shortcomings, the grid has already reached a level at which it can be used for real analysis. We are therefore in the process of making it available on the internet, together with the tools that 
we can currently offer to use it. It will be freely available for any interested researcher by the end of 2020. 

Finally, we would like to justify our choices of parameters and comment on future expansions of the grid. It might rightfully be argued that we ignore some important parameters in the grid. The most consequential omission is the turbulent velocity because it affects the analysis of optical He lines. Instead of exploring these effects, we simply use a generic  V$_{\rm turb}$=20 km s$^{-1}$ throughout the wind. Furthermore, we should not really use the same ionic species for the hottest O-type and for the B-type stars because the ionization level of elements can be different for these stars (the most notable example is Fe). However, we have limited access to the computational resources, and compromises and hard choices have to be made. With our selection of parameter space, we intend to create a grid that is sophisticated enough to be useful in the analysis of observed spectra, but the computational requirements of its production stay within the limits. It is not a trivial task to achieve this compromise because the limits can easily be exceeded by introducing a seemingly minor adjustment in the parameter space.  

Nevertheless, if there is future demand for an expansion of the parameter space (e.g., have a range in turbulent velocity), we will address it with a follow-up project. The grid can be  augmented in the future, and not
only by us, if sufficient computing resources are available. Furthermore, there are ways to improve our grid that do not involve running more models on supercomputers. We can, for example, appeal to the users of CMFGEN to donate their no longer needed models to the grid. We suspect that there are hundreds or even thousands of models sitting idle in computers around the world. Including such a diverse set of models into our grid will be challenging, but might be feasible.  
 
We describe our choice of parameter space in the context of the intended purpose of the grid. We wish to provide a tool for easy  and rapid  analysis of the stellar spectra that will serve well a variety of special applications where the need for rapid analysis outweighs the need for high-accuracy parameter determination. Examples of such applications include statistical analyses of large stellar samples or population synthesis models. However, we do not mean to replace the traditional modeling by using only the grid for a  detailed analysis of a particular star. In such cases, the investigators still have to fine-tune the parameters by running models with the code of their preference (which does not have to be CMFGEN) to complete the task. Using the grid will speed up the analysis by eliminating the need to run numerically costly seed models in the early phases of the investigation, and will provide good initial conditions for the fine-tuning phase. The time saving is very significant. For example, our reanalysis of $\epsilon$ Ori lasted only about two days, while the duration of the original work by \cite{pue16} was several weeks. We are talking about a five- to tenfold decrease in the required time.  
   
\section{Simple test to demonstrate the usefulness of our grid}
\label{sec:simtest}

We demonstrate the benefits that our grid offers by reanalyzing $\epsilon$~Ori. This B0~Ia 
supergiant was recently studied by \cite{pue16} in the traditional way, that is, by producing every 
model that was needed for the analysis. Some members of our team were part of the group that 
performed this study, and so they had access to the observational data used in the work of 
\cite{pue16}. It is therefore convenient and very useful to see the result we would obtain by 
using the models available in our grid and without running CMFGEN at all. Before presenting our  
results, we have to mention some limitations that our reanalysis has in comparison with the original 
study. For example, \cite{pue16} used the experimental version of CMFGEN and included high-resolution 
X-ray observations in the analysis to augment the UV and optical spectra. Because we are used the 
commonly available version of CMFGEN, we cannot reproduce their results in the X-ray range and so we 
cannot verify the additional information gained by using this spectral range (e.g., additional constraints 
on the mass-loss rate and abundances). Furthermore, our models assume vacuum in the interclump medium, as 
opposed to theirs, which included tenuous but not empty regions between the clumps. By doing so, they 
successfully reproduced the FUV \ion{O}{VI} and \ion{N}{V} lines, which we were not able to do. 
Nevertheless, in the following sections we demonstrate that we can reproduce their results quite 
well and that process in a spectral analysis can be made by using our grid and before running any 
simulation.

Finally, we note that because we used the very same observations in our analysis as \cite{pue16} did, 
we omit the description of the data reduction and the observations in this paper. We refer to the relevant sections of \cite{pue16} for more information on this subject.

\subsection{$\epsilon$~Ori (\object{HD37128})}

This supergiant star is the central star of the Orion belt and belongs to the Orion OB1 association. 
It has been studied many times in the past. The most recent works include \cite{kud99}, who 
found \Teff$=28,000$~K and \logg$=3.00$ for this star using the code FASTWIND, as well as a series 
of CMFGEN studies by \citet{cro06}, \citet{sea08}, and \cite{pue16}. The first two authors found 
very similar effective temperatures, that is, 27,000 and 27,500~K, respectively, but they diverged 
significantly in the value of \logg\ (2.9 and 3.1, respectively). Previous studies also reported 
a wide range of mass-loss rates, but in general, lower than $5.0\times 10^{-6}$ $M_{\odot}$ yr$^{-1}$. 
The majority of these studies assumed smooth wind (no clumping).

However, the most relevant results for our analysis are those found by \cite{pue16}. They reported 
\Teff$=27,000\pm 500$~K, \logg$=3.00\pm 0.05$, a mass-loss rate of \Mdot$\sim 10^{-7}$ $M_{\odot}$ 
yr$^{-1}$, $V_{\infty}\sim 1800$ km s$^{-1}$, and a highly clumped and slowly accelerating wind 
($F_{cl}=0.01$, $\beta >2.0$) for $\epsilon$~Ori. However, their volume filling factor is not really 
well defined because of uncertainties in considering the degree to which the UV lines are affected 
by the velocity-space porosity. If they are not affected too much, the wind has to be very clumped 
and the mass-loss rate has to be low in order to fit the \ion{S}{IV} $\lambda\lambda$1062,1073 and 
the \ion{P}{V} $\lambda\lambda$1118,1128 profiles. Because of this uncertainty, the authors considered four 
variations of the same model with \Fcl$=0.01$, 0.05, 0.1, and 1.0 as their best models, each with 
mass-loss rates adjusted to conserve the ratio \Mdot$/\sqrt{F_{\rm cl}}\sim 1.6\times 10^{-6}$ 
$M_{\odot}$ yr$^{-1}$. We refer to their Tables 4 and 5 to learn more about 
their best results.
%
\begin{table}
\begin{center}
\caption{Weights used by FIT\textit{spec} for $\epsilon$~Ori.
\label{tab:fitspecData}}
\begin{tabular}{llcc}
 & \multicolumn{1}{c}{Lines or ratios} &\multicolumn{1}{c}{Initial weights} &
\multicolumn{1}{c}{Final weights}\\
\hline
\multicolumn{4}{l}{ }\\
\multicolumn{4}{l}{\Teff \hspace{0.1cm} fit:}\\
 & \ion{He}{II}  4541/\ion{He}{I} 4471 & 0.25 & 0.20\\
 & \ion{He}{II}  4200/\ion{He}{I} 4026 & 0.25 & 0.30\\
 & \ion{He}{II}  4200/\ion{He}{I} 4144 & 0.25 & 0.30\\
 & \ion{He}{II}  4541/\ion{He}{I} 4387 & 0.25 & 0.20\\
\multicolumn{4}{l}{ }\\
\multicolumn{4}{l}{\logg \hspace{0.1cm} fit:}\\
 & H I 3835 & 0.17 & 0.17 \\
 & H I 3889 & 0.17 & 0.17 \\
 & H I 3970 & 0.17 & 0.17 \\
 & H I 4102 & 0.17 & 0.17 \\
 & H I 4341 & 0.17 & 0.16 \\
 & H I 4861 & 0.17 & 0.16 \\
\multicolumn{4}{l}{ }\\
\multicolumn{4}{l}{Combined fit:}\\
 & Ratios \ion{He}{II}/\ion{He}{I} & 0.50 & 0.50 \\
 & EWs \ion{H}{I} & 0.50 & 0.50 \\
\hline
\end{tabular}
\end{center}
\end{table}

\subsection{Analysis by FIT\textit{spec}}
\label{sec: FITspec}

As previously mentioned, we intend to provide tools that aid the comparison of the model and observed spectra and accelerate the search for the best-fitting models. FIT\textit{spec}, the first of these tools, is already operational, therefore it is natural to start our reanalysis of $\epsilon$~Ori by applying this code.  To start an analysis by FIT\textit{spec}, the user must measure the equivalent widths (EWs) in the observations of the $\lambda\lambda$3835, 3889, 3970, 4101, 4349, 4861 lines of the H Balmer series, and those of the \ion{He}{II} $\lambda\lambda$ 4541 and 4200, as well as the \ion{He}{I} $\lambda\lambda$ 4471, 4387, 4144, and the \ion{He}{I}+\ion{He}{II} blend at $\lambda$4026; and provide them as inputs. The program then compare these EWs with those measured in the models and finds the closest match. This procedure is relatively fast because FIT\textit{spec} has a complete library of the EWs for these lines measured in each model spectra of the grid. Instrumental and rotational broadening of the observed spectra are not of concern because they do not affect the measured EWs. 

FIT\textit{spec} uses the EW of the Balmer lines and the EW ratios of \ion{He}{II} and \ion{He}{I} lines to search for the best-fitting models in the \logg\ -- \Teff\ space. The strategy is the following. First, the code calculates the ratios of 
\begin{equation}
\label{e:He1}
\frac{{\rm EW(\ion{He}{II} \lambda4541)}}{{\rm EW(\ion{He}{I} \lambda4471)}},
\end{equation}
\begin{equation}
\label{e:He2}
\frac{{\rm EW(\ion{He}{II} \lambda4200)}}{{\rm EW(\ion{He}{I} \lambda4026)}},
\end{equation}
\begin{equation}
\label{e:He3}
\frac{{\rm EW(\ion{He}{II} \lambda4200)}}{{\rm EW(\ion{He}{I} \lambda4144)}},
\end{equation}
\begin{equation}
\label{e:He4}
\frac{{\rm EW(\ion{He}{II} \lambda4541)}}{{\rm EW(\ion{He}{I} \lambda4387)}},
\end{equation}
then the relative differences, or as we call them, the  {\it "\textup{errors}"}, are calculated by
\begin{equation}
{\rm Error}_{\rm EW}=\frac{{\rm EW}_{\rm obs}-{\rm EW}_{\rm model}}{{\rm EW}_{\rm obs}}.
\end{equation}
for each H Balmer line, and by
\begin{equation}
{\rm Error}\left(\frac{\ion{He}{II}}{\ion{He}{I}}\right) =
\frac{\left(\frac{\ion{He}{II}}{\ion{He}{I}}\right)_{\rm obs} -
\left(\frac{\ion{He}{II}}{\ion{He}{I}}\right)_{\rm model}}
{\left(\frac{\ion{He}{II}}{\ion{He}{I}}\right)_{\rm obs}}.
\end{equation}
for each \ion{He}{II} to \ion{He}{I} ratios defined by Eqs.~(\ref{e:He1})--(\ref{e:He4}).

FIT\textit{spec} then calculates the weighted averages of these errors, 
where the weights need to be specified by the user a priori. Weights are usually assigned to lines 
that reflect the quality of their signal-to-noise ratio (S/N) (see Table~\ref{tab:fitspecData} for the weights we used for $\epsilon$~Ori). 
Then, models with average errors smaller than 30\% in either the EWs of the H Balmer lines or the EW ratios of the He lines are selected, and among these models FIT\textit{spec} identifies the model that has the smallest total error,
\begin{equation}
{\rm Error}_{\rm tot}=\sqrt{ \sum \left({\rm Error}_{\rm EW}\right)^{2} + 
\sum \left[{\rm Error}\left(\frac{\ion{He}{II}}{\ion{He}{I}}\right)\right]^{2}},\end{equation}
where the summation is over all H lines and He ratios. Then, the user has the option to adjust the weights, if necessary, and restart the procedure. After a few iterations, the best-fitting models are found. More about the code FIT\textit{spec} and the way it works can be found in \cite{fie18}.

After running FIT\textit{spec}, we found that models with \Teff=27,320$\pm$1500~K and \logg=3.27$\pm$0.29 fit the observations best that agree very well with the results of previous studies.

\begin{table}
\begin{center}
\caption{Models in the grid that best fit the observations of $\epsilon$~Ori 
\label{tab:parameters}}
\begin{tabular}{lccc}
Parameter                        &   Model 1       &       Model 2   &   Model 3 \\ \hline
\Teff  \hspace{0.1 cm}(K) &     26,540        &       26,540    &     26,980 \\
L   (10$^{5}$L$_{\odot}$)  &       4.196       &        4.196     &       4.491 \\
\logg                           &      3.02          &         3.02       &         3.06 \\
R  (R$_{\odot}$)               &    30.62          &        30.62      &       30.65 \\
\Mdot\   (10$^{-7}$\Msun yr$^{-1}$) &           1.526   &            3.737  &     10.82  \\
\Vinf\  (km s$^{-1}$)          &     1,414          &     1,414         &        1,473 \\
$\beta$                         &        2.0           &        2.0          &         2.0    \\
\Fcl\                                  &               0.05  &                0.3  &          0.6    \\ 
\vsini\  (km s$^{-1}$)         &         80            &       80            &          80     \\
                                            \multicolumn{4}{c}{ }                                       \\
M$_{\rm ini}$$^a$ (\Msun)      &        40      &       40            &           44    \\
Age$^a$ (10$^6$yr)                &      4.47     &      4.47          &         3.98   \\
X(He)$^a$                               &   0.266      &    0.266          &        0.266  \\
X(C)$^a$                                 &   2.31E-3  &    2.31E-3       &    2.31E-3   \\
X(N)$^a$                                 &   6.59E-4  &    6.59E-4       &    6.59E-4   \\
X(O)$^a$                                 &   5.73E-3  &    5.73E-3       &    5. 74E-3  \\
Chemical comp.$^a$               & \multicolumn{3}{r}{Solar without rotational enhancement} \\ \hline
\end{tabular}
\end{center}
 $^a$From the evolutionary models of \cite{eks12}.
\end{table}

\begin{figure*}
\centering
\includegraphics[width=0.9\textwidth]{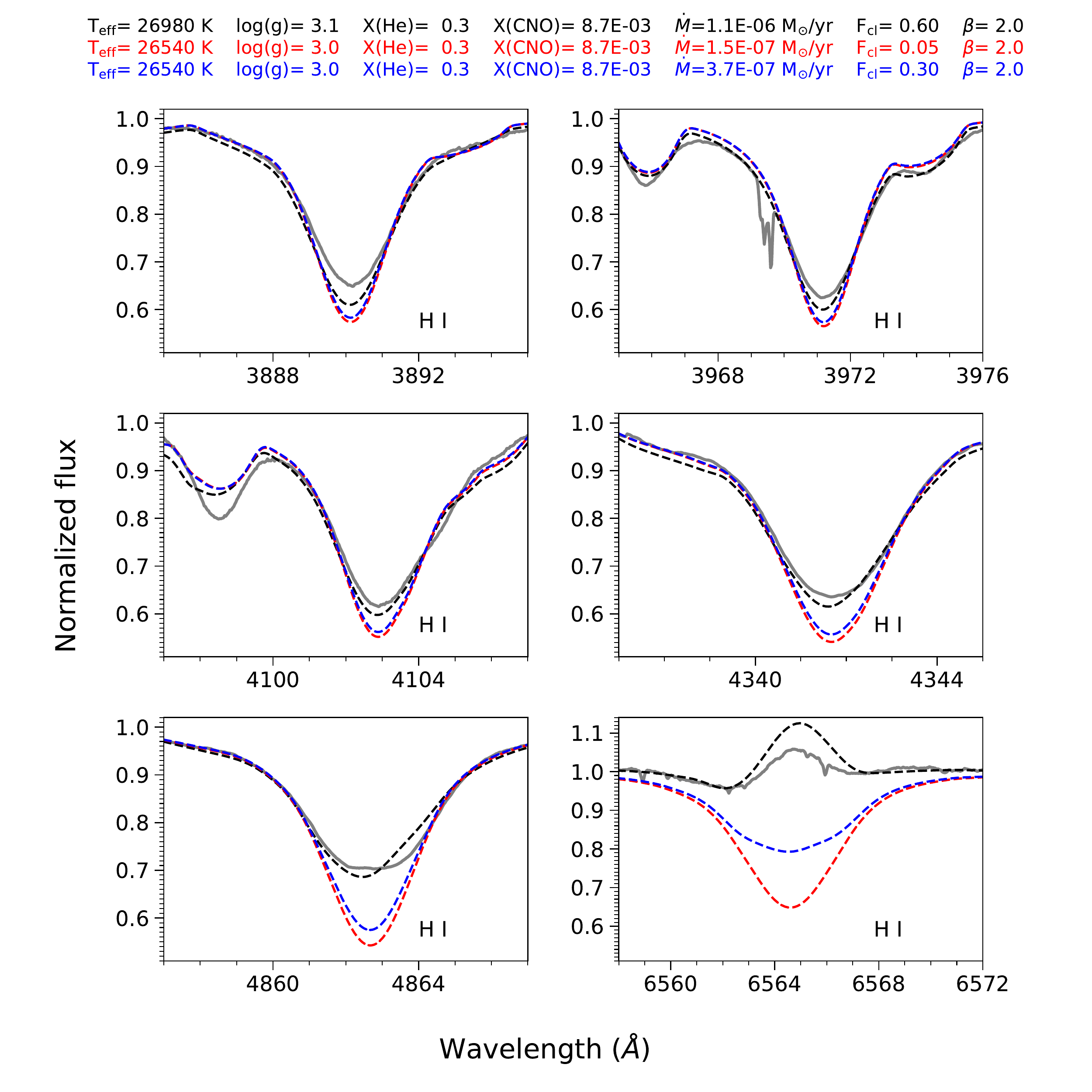}
\caption{Comparison of our best-fitting models (dashed black, red, and blue lines) with the observed \ion{H}{I} lines for $\epsilon$~Ori (solid gray line). The relevant model parameters are color-coded above the panels.}
\label{f:H_best}
\end{figure*}

\begin{figure*}
\centering
\includegraphics[width=1.0\textwidth]{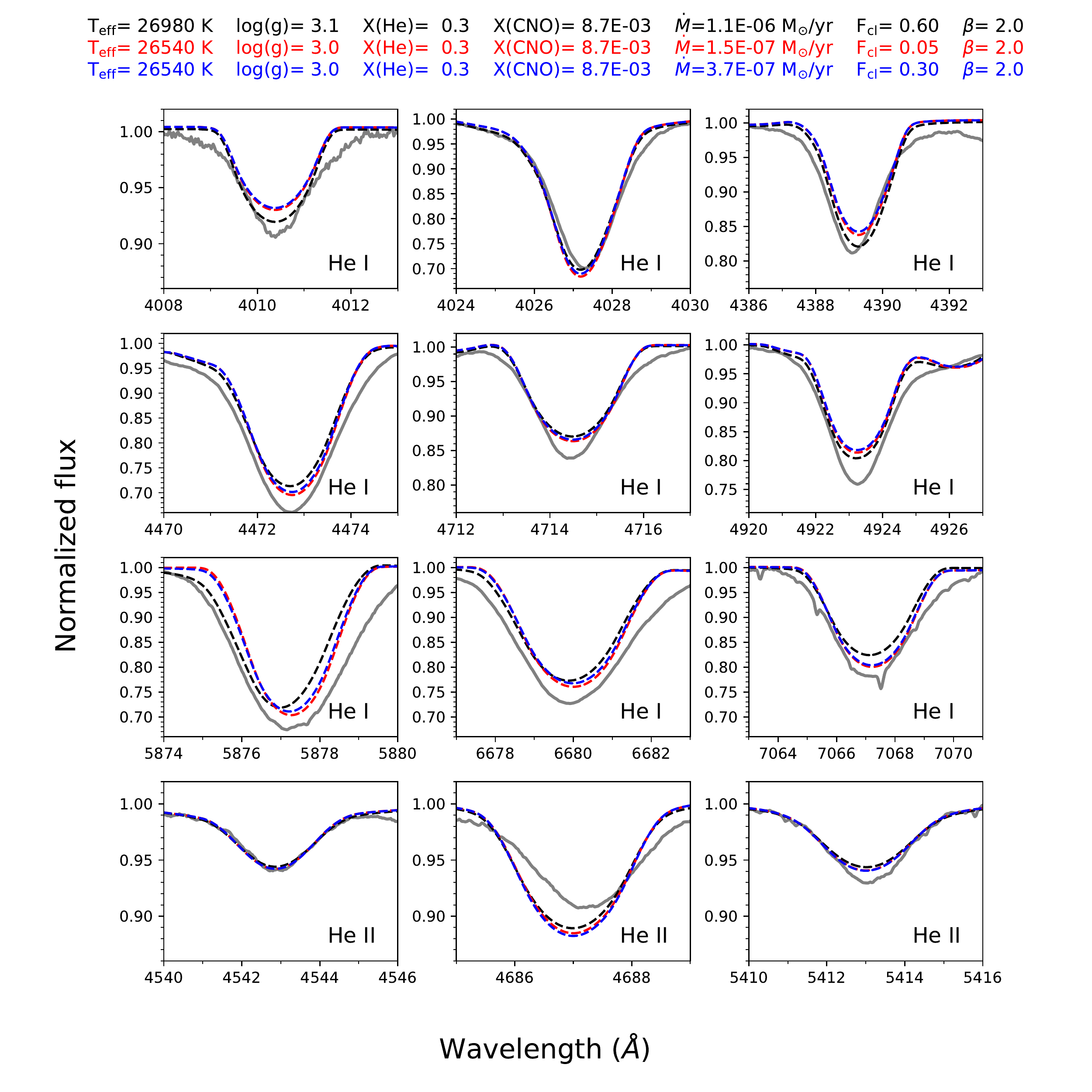}
\caption{Same as Fig.~\ref{f:H_best}, but the comparison is for the \ion{He}{I} and \ion{He}{II} lines.}
\label{f:He_best}
\end{figure*}

\begin{figure*}
\centering
\includegraphics[width=0.8\textwidth]{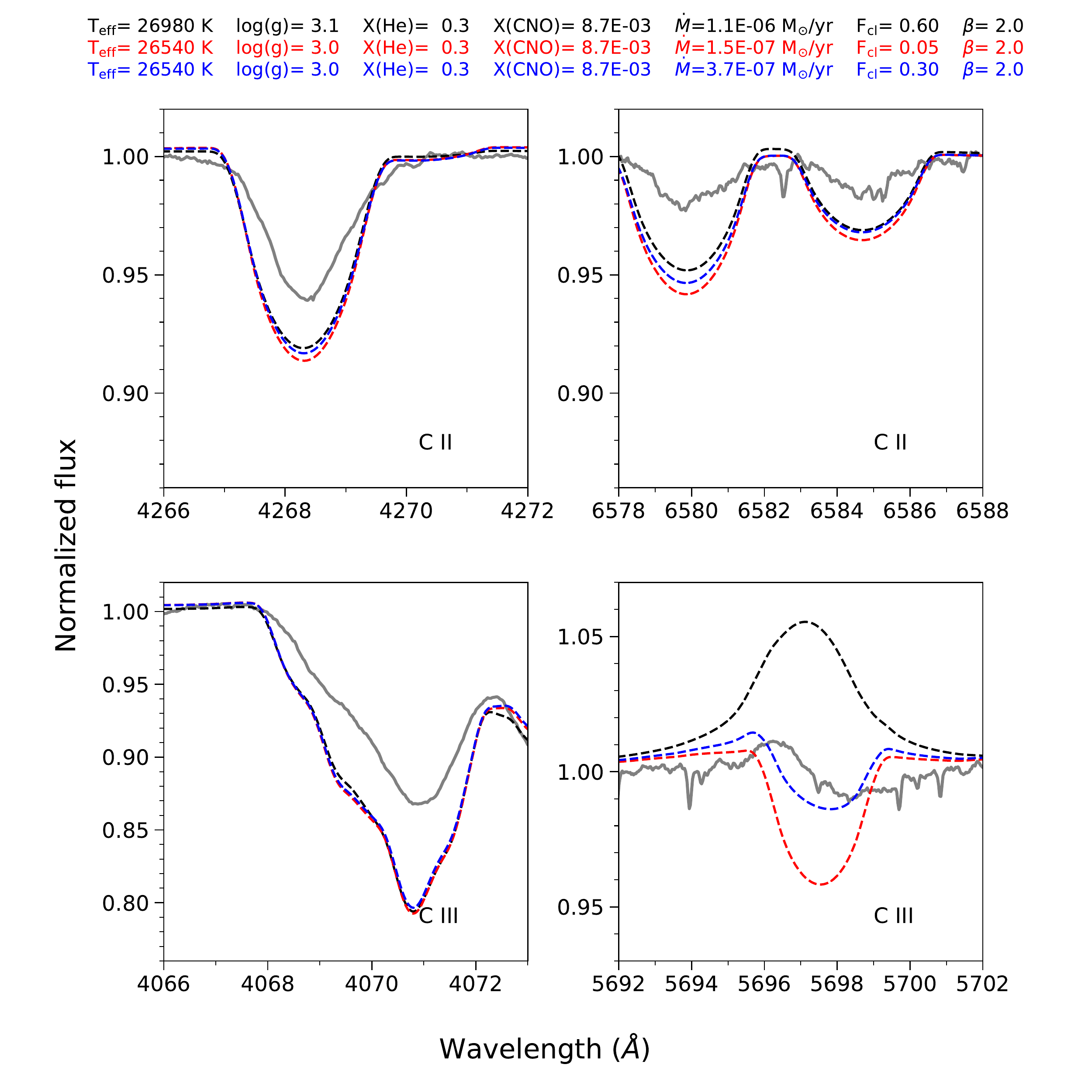}
\caption{Same as Fig.~\ref{f:H_best}, but the comparison is for selected C lines.}
\label{f:C_best}
\end{figure*}

\begin{figure*}
\centering
\includegraphics[width=0.8\textwidth]{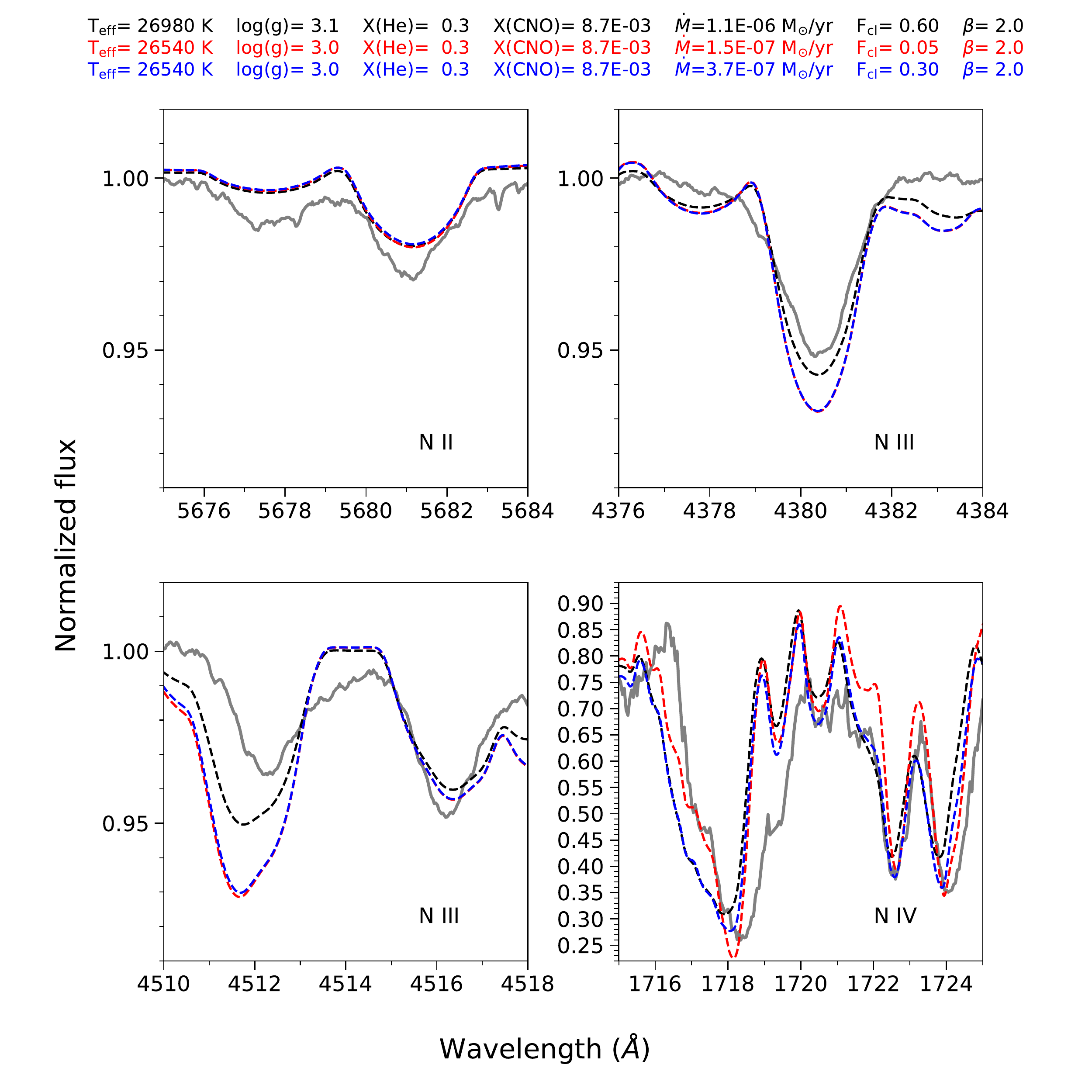}
\caption{Same as Fig.~\ref{f:H_best}, but the comparison is for selected N lines.}
\label{f:N_best}
\end{figure*}

\begin{figure*}
\centering
\includegraphics[width=1.0\textwidth]{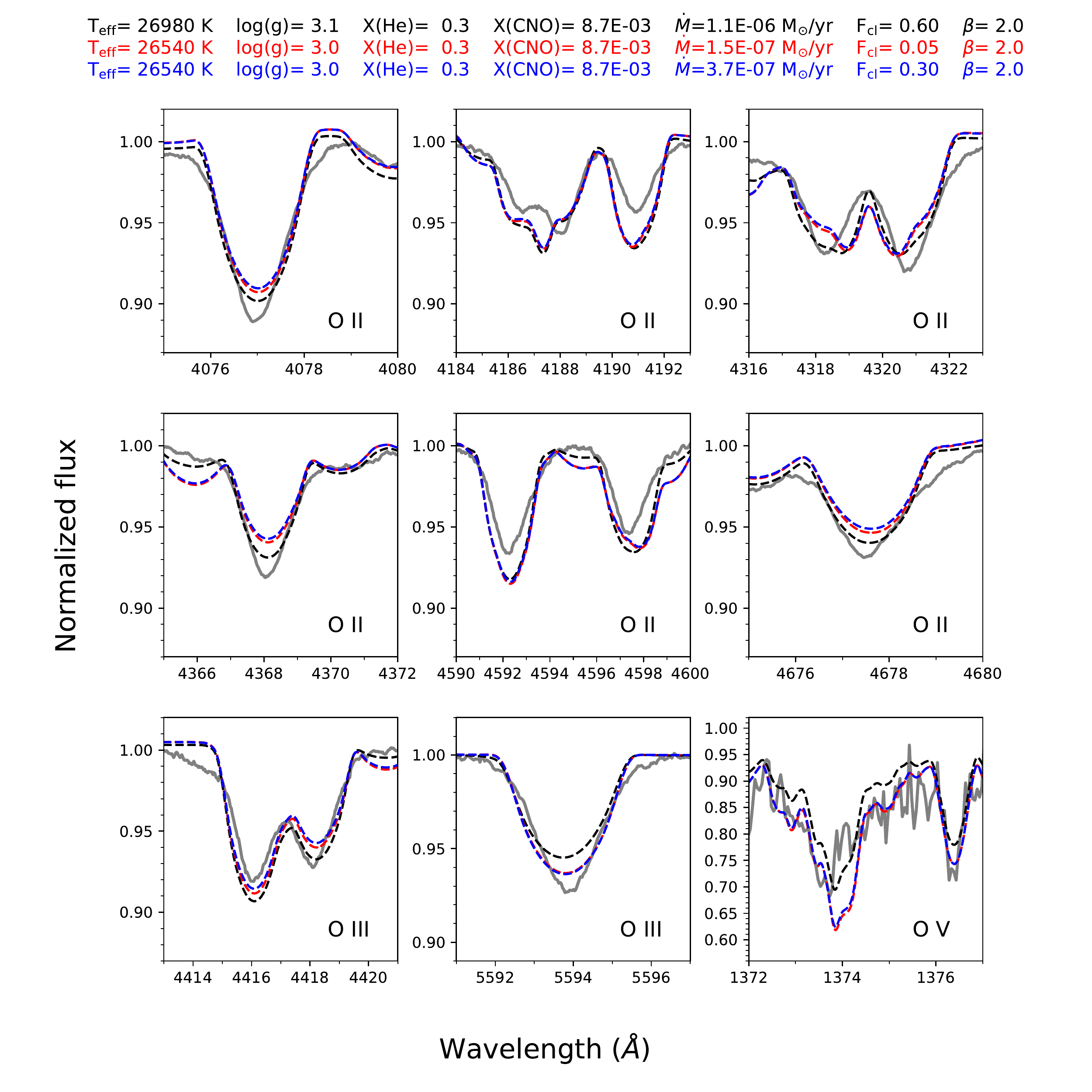}
\caption{Same as Fig.~\ref{f:H_best}, but the comparison is for selected O lines.}
\label{f:O_best}
\end{figure*}

\begin{figure*}
\centering
\includegraphics[width=0.8\textwidth]{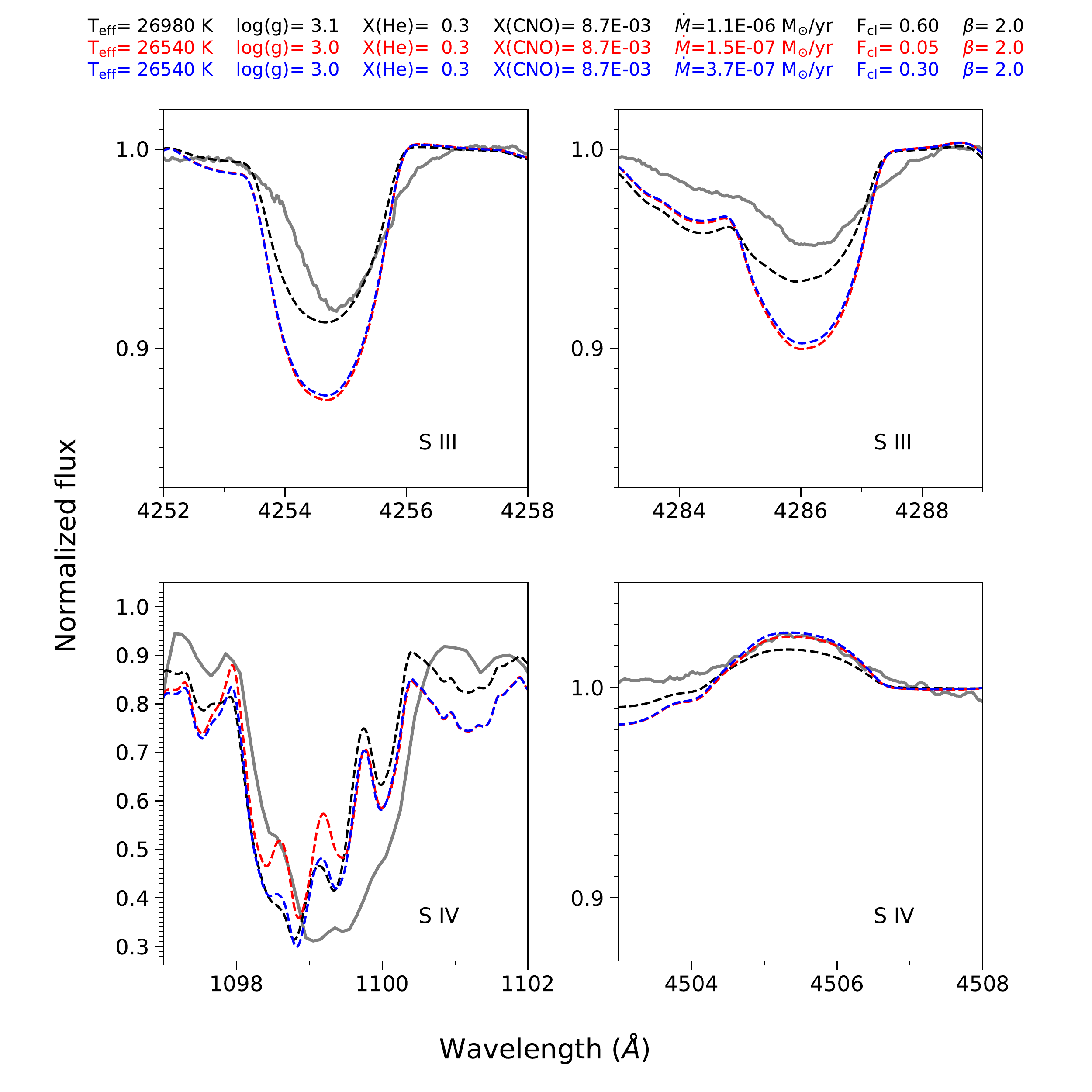}
\caption{Same as Fig.~\ref{f:H_best}, but the comparison is for selected S lines.}
\label{f:S_best}
\end{figure*}

\begin{figure*}
\centering
\includegraphics[width=0.9\textwidth]{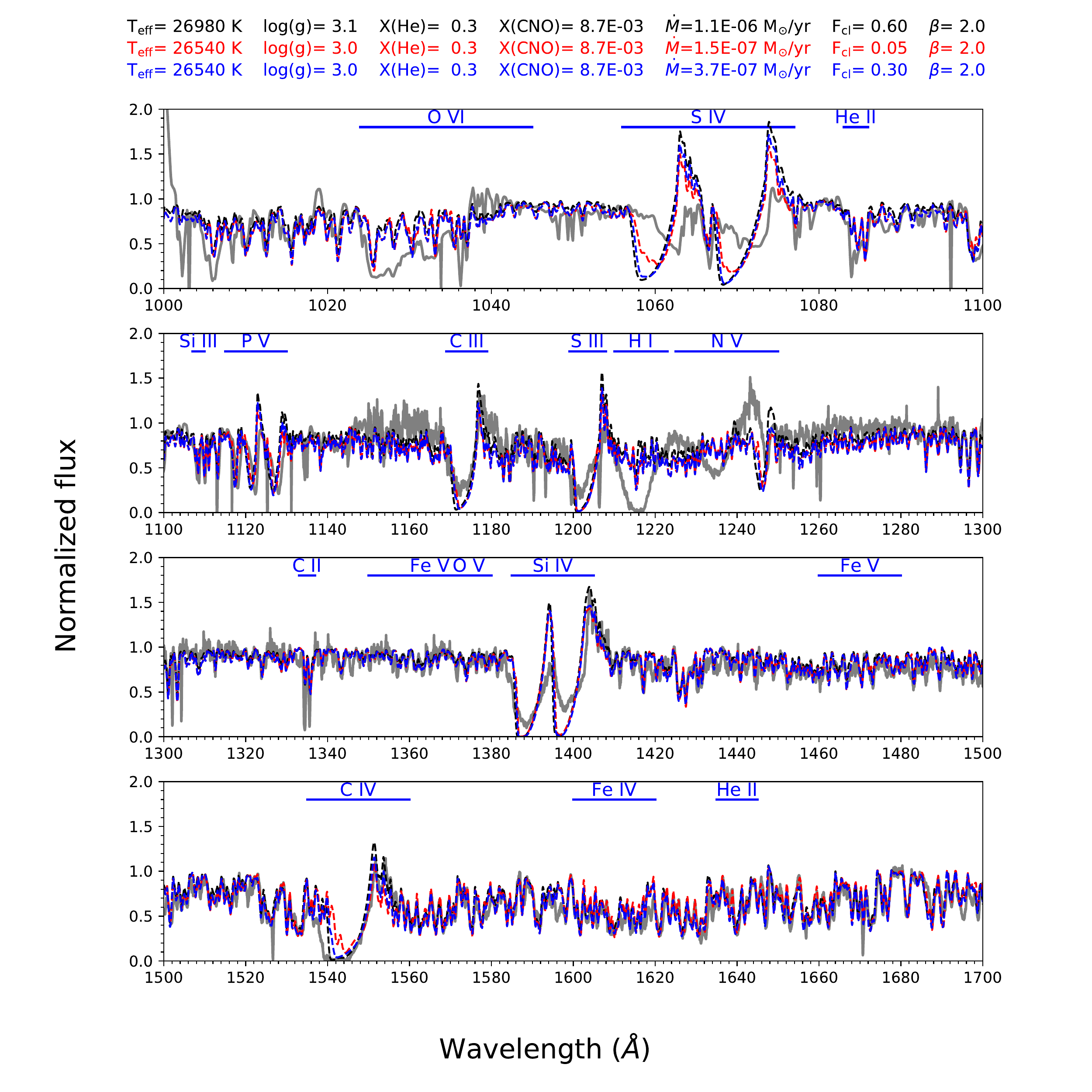}
\caption{Comparison of our best-fitting models (dashed black, red, and blue lines) with the UV spectra observed for $\epsilon$~Ori (solid gray line). The relevant model parameters are color-coded above the panels, and important UV lines marked above the spectra.}
\label{f:UV_best}
\end{figure*}

\subsection{Visual inspection of the spectra} \label{sec:visual}

After identifying the relevant parameter ranges in \Teff\ and \logg\ by FIT\textit{spec}, 
we visually inspected the models within these ranges. This was done with another tool that we created to help in the search for best-fitting models. The tool allows selecting specific lines from any wavelength region for display on the screen and compare the observations with selected models. The display is very similar to Fig.~\ref{f:He_best}, for example. Several of our figures are simple snapshots of this display. Then, we can browse through a number of models and visually inspect the fit. The user has the option of changing the parameter ranges of the models that are queued for inspection, to apply rotational or instrumental broadening to the model spectra, to Doppler-shift them, zoom in and out, and the displayed lines can also be changed. The comparison can be made in the form of normalized or calibrated flux spectra.  Unfortunately, the interaction with the code is not yet user-friendly, therefore we have not yet released it. However, we are developing a graphic user interface to make it user friendly. When this code is optimized and easy to apply, it will be released and will be a powerful tool if used in combination with FIT\textit{spec}. 

With the help of FIT\textit{spec} we were able to decrease the number of models that are to be inspected visually to a few hundred, and then applied a rotational broadening with \vsini= 80 km~s$^{-1}$. The convolution of the model spectra with a broadening profile adequately accounts for the rotation and macro-turbulence for slowly rotating stars \citep{hil12}.  Wind-free line profiles were used to estimate the projected rotational velocity \vsini. After several hours of inspection, we then found that 26,500~K<\Teff<27,000~K, 3.0<\logg<3.1,  $\beta$=2.0$\pm$0.3, and \Vinf$\ge$1500 km s$^{-1}$ are the best values for the stellar and wind 
parameters (see the parameters of the best-fitting models in Table~\ref{tab:parameters}). The error estimates are based on the ranges of acceptable models and on the parameter resolutions that we have in the grid. 
Unfortunately, we were only able to derive a lower limit for \Vinf\ because the coverage that our grid 
still has in this parameter is limited. For the mass-loss rates we found that \Mdot\ could vary between 
$1.5\times10^{-7}$ \Msun yr$^{-1}$ and $1.1\times10^{-6}$ \Msun yr$^{-1}$, depending on 
the adopted \Fcl\ so that \Mdot$/\sqrt{F_{\rm cl}}\sim1.4\times10^{-6}$ \Msun yr$^{-1}$ is 
approximately conserved. Our analysis mildly favors the combination of \Mdot=$1.1\times10^{-6}$ 
\Msun yr$^{-1}$ with \Fcl=0.6 and indirectly suggests that vorosity highly affects the UV 
resonance lines (see the discussion in \S\ref{sec:disc}). These values agree well with the 
results of previous studies. Figs.~\ref{f:H_best}--\ref{f:S_best} show our best-fitting models with the observed spectra for important H, He, C, N, O, and S lines. In the following, we discuss how the different parameters affect the spectral diagnostics and describe our results. 

Fig.~\ref{f:H_best} shows how well  our models fit the Balmer series of hydrogen, especially the model with the highest mass-loss rate (Model 3 in Table~\ref{tab:parameters} and the dashed black lines in Fig.~\ref{f:H_best}). The Balmer series is normally used to measure \logg, but the figure reveals that other parameters also affect the H lines. For example, there is 
a significant effect due to mass loss as the absorption profiles of H$\alpha$ and H$\beta$ are filled in 
by the emission from the wind. We also observe significant differences in the synthetic profiles for different values of \Fcl, especially in the case of H$\alpha$. However, the line profiles should not vary with \Fcl\ if the mass-loss rate is adjusted to conserve \Mdot$/\sqrt{F_{\rm cl}}$. As we explain in \S~\ref{sec:disc}, because of the way in which CMFGEN introduces clumping (see Eq.~(\ref{e:clumping})) and the slow acceleration of the wind of $\epsilon$~Ori, an extended transition zone exists between the smooth wind near the stellar surface and the outer region with constant \Fcl. This transition zone has a profound effect on the profiles of the lower order Balmer lines. Fortunately, the effects of the wind emission are negligible on the higher order Balmer lines, as Fig.~\ref{f:H_best} clearly shows, so we can use them to measure \logg. The good fit between the models and the observations in the wings of these lines (see the top panels of Fig.~\ref{f:H_best}) suggests that the most probable value of \logg\ is about 3.

Fig.~\ref{f:He_best} shows the comparison of our best-fitting models with the optical \ion{He}{I} 
and \ion{He}{II} lines observed for $\epsilon$~Ori. Although the \ion{He}{II} lines are very weak for this star, the comparison clearly indicates that \Teff\ has to be about 27,000~K. No other parameter
significantly affects the He lines, at least in the parameter space that we use in our grid. The derived \Teff\ is also consistent with the observed C, N, O, and S lines, as illustrated by Figs.~\ref{f:C_best}--\ref{f:S_best}. These figures show lines that originate from consecutive ionization levels and do not suggest that \Teff\ needs to be revised. The fit is surprisingly good considering the limited resolution and covering that our grid has in the parameter space. Figs.~\ref{f:N_best}--\ref{f:S_best} also indicate that the abundances of N, O, and S are about right in the best-fitting models, while Fig.~\ref{f:C_best} suggests that we should lower the carbon abundance  by a factor of few. In this figure all synthetic profiles are stronger than the observations, regardless of the level of ionization. This conclusion agrees very well with the findings of \cite{pue16}, who needed a C abundance a factor of 2 to 4 lower than those in our models to fit these lines. However, the objective of this paper is to show how closely we can match the right stellar and wind parameters by using only the grid, therefore we did not adjust the carbon abundance.   

The most useful spectral region to estimate the mass-loss rate \Mdot\ and terminal velocity \Vinf\ 
is the UV region. Here, we encounter strong resonance lines of the dominant ionization states 
in the winds of massive stars. These resonance lines often show P-Cygni profiles, which are useful for 
measuring \Mdot\ and \Vinf\ (see, e.g., the \ion{C}{IV} doublet around 1550 \AA\ or the \ion{Si}{IV} 
doublet around 1400 \AA\ in Fig.~\ref{f:UV_best}). Unfortunately, we cannot reproduce many of these 
lines well for various reasons. For example, we have a problem to fit the \ion{N}{V} doublet around 1240 
\AA\ because our models do not include interclump medium. The UV region is also not 
useful for estimating the $\beta$-parameter and \Fcl\ because most of the P-Cygni profiles are saturated (and probably strongly affected by vorosity). To derive mass-loss rates, the UV spectra show a somewhat contradictory situation. While the \ion{Si}{IV} $\lambda\lambda$1394, 1403 doublet suggests low values of \Mdot\ to fit the \ion{C}{IV} $\lambda\lambda$1548, 1550 we would need much higher mass-loss rates. However, \Mdot$>1.5\times 10^{-6}$ \Msun yr$^{-1}$ would result in H$\alpha$ emission, which is not observed. We therefore conclude that $1.1\times 10^{-6}$ \Msun yr$^{-1}>$\Mdot$>10^{-7}$ \Msun yr$^{-1}$ is 
the best estimate we can have. The actual value depends on the adopted \Fcl.  Fig.~\ref{f:UV_best} also shows that the terminal velocity has to be higher than 1,500 km s$^{-1}$, but our grid 
does not yet have the necessary coverage in \Vinf\ to determine its exact value. 

\section{Discussion}
\label{sec:disc}

In Figs.~\ref{f:A1} and \ref{f:A2} we show the general comparisons of our best models (dashed black, red, 
and blue lines) with the observations (gray lines). With these figures we intended to reproduce 
Figs. A1--A3 in the appendix of \cite{pue16} for an easy comparison of our and their results. These
figures also show that the overall fit in the optical range is very good, especially for the model with 
\Mdot$=1.1\times 10^{-6}$ $M_{\odot}$ yr$^{-1}$ and \Fcl$=0.6$. However, as was mentioned earlier, 
there are significant discrepancies in the UV.

The top panel of Fig.~\ref{f:UV_best} shows that our models lack the \ion{O}{VI} profile around 1032~\AA,\, 
which is expected because our models do not take the interclump medium into account. A graver problem is 
that \ion{S}{IV} around 1070 \AA\ is too strong for our models, especially for \Mdot=$1.1\times 10^{-6}$ $M_{\odot}$ 
yr$^{-1}$ and \Fcl$=0.6$. The same problem led \cite{pue16} to the conclusion that the wind is extremely 
clumped (\Fcl$\sim 0.01$) and \Mdot\ is low. However, they have already raised the possibility that the 
fact that CMFGEN does not yet take the velocity-space porosity into account might be responsible for 
the anomalously strong \ion{S}{IV} profiles in the models. Since the publication of their paper, several 
studies have concluded that the UV P-Cygni profiles should indeed be affected strongly by the vorosity 
effects. For example, one of the main conclusion of \cite{sun18} was that velocity-space porosity is 
critical (in their words) for the analysis of UV resonance lines in O stars. Taking these 
effects into account might fix our problems because it would weaken these lines for the same \Mdot. In 
short, the theory of velocity-space porosity assumes that as the material is swept up in dense clumps, the 
process also creates gaps in the velocity distribution of the material. These gaps then allow the escape 
of radiation that otherwise would be absorbed in smooth wind.

A possible other solution for the weak P-Cygni profiles in the observations could be the low abundances of 
the species in question. This possibility was raised and quickly dismissed when the same problem was encountered with the \ion{P}{V} $\lambda\lambda$1118,1128 lines in the spectral analysis of early-O stars. Although we did not adjust abundances in our work, there is nothing in other spectral regions 
that would suggest anomalously low abundances. The sulphur lines shown in Fig.~\ref{f:S_best} suggest that the sulphur abundance is more or less correct in our models.

\begin{figure*}
\centering
\includegraphics[width=0.33\textwidth]{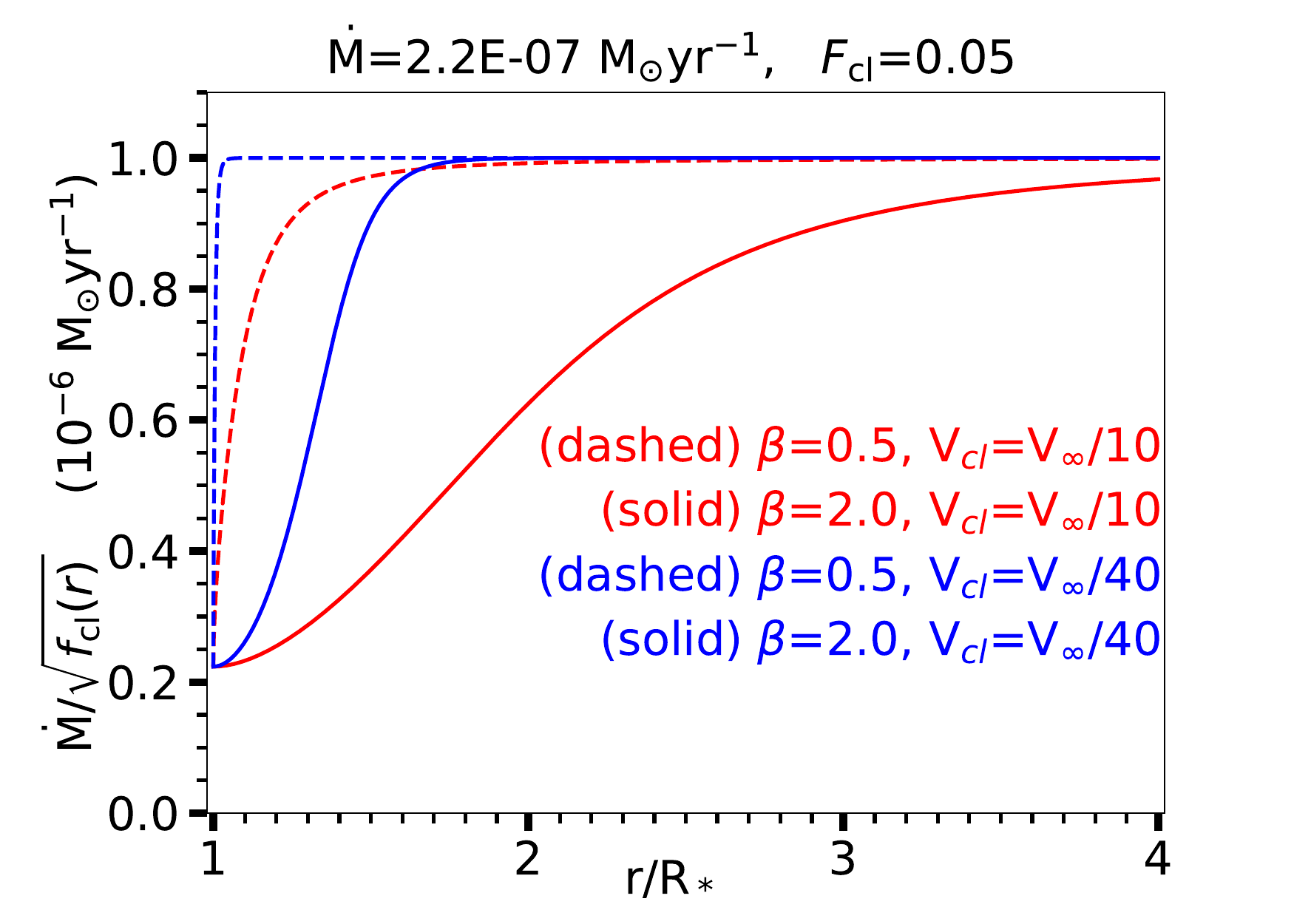}
\includegraphics[width=0.33\textwidth]{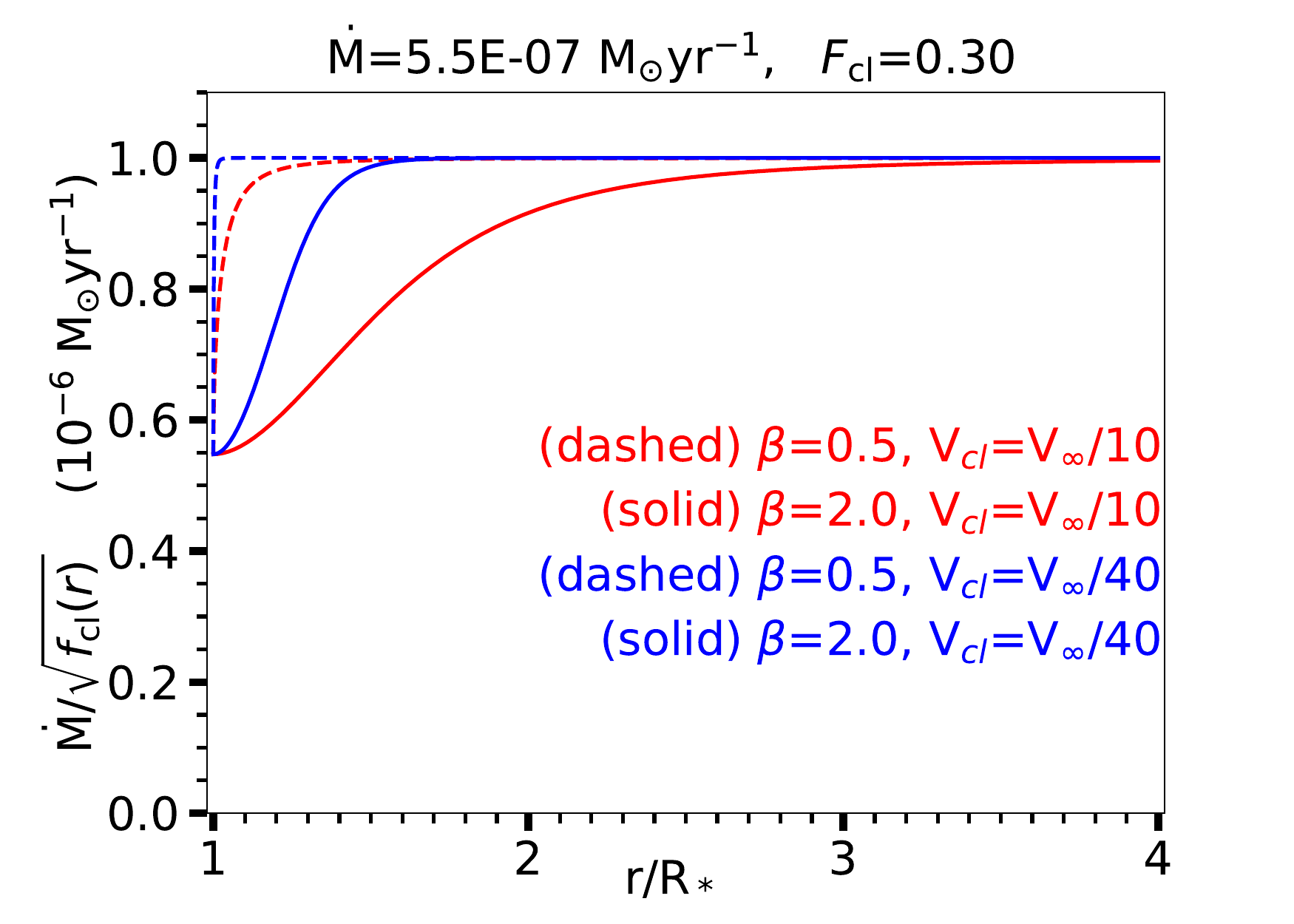}
\includegraphics[width=0.33\textwidth]{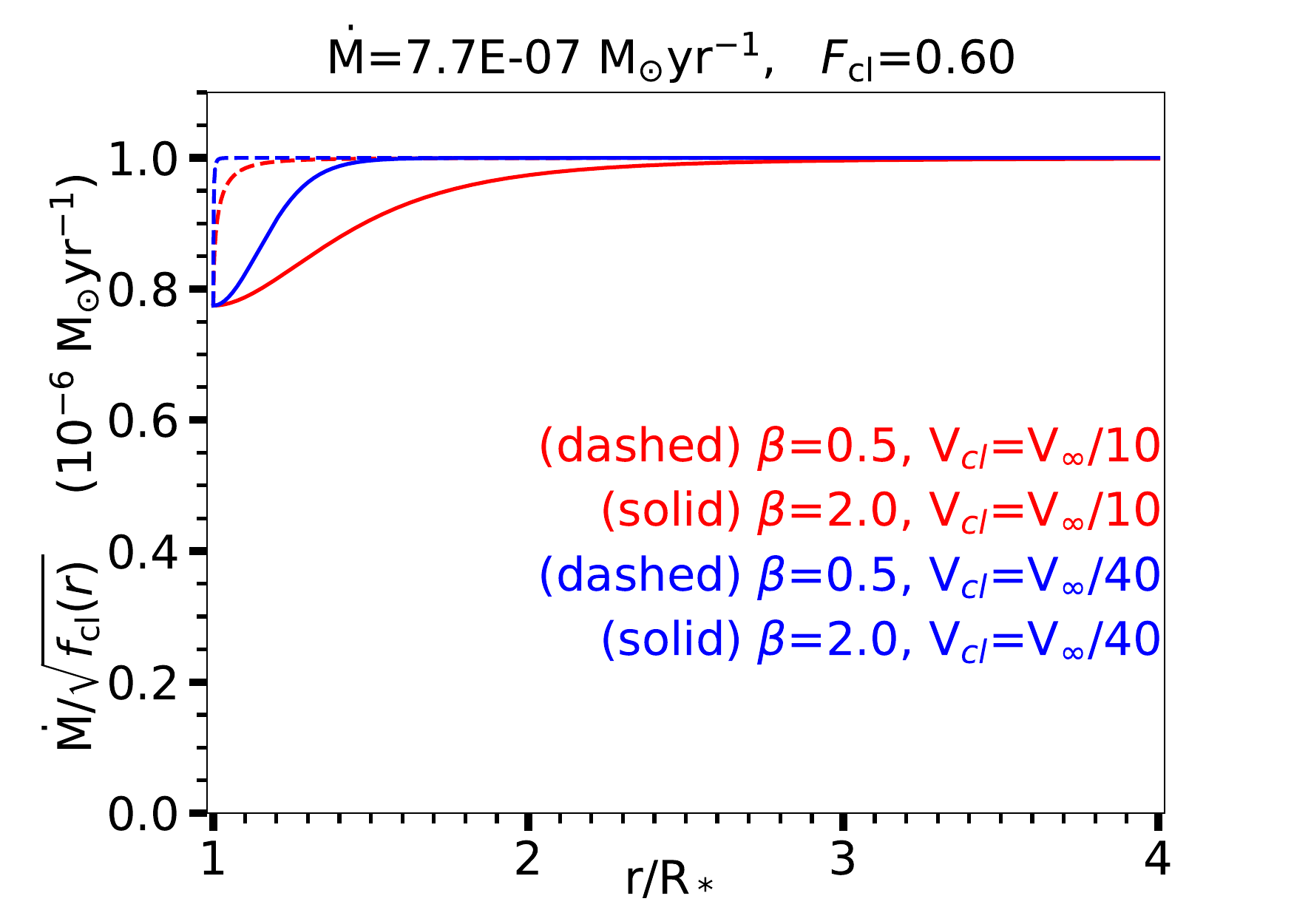}
\caption{True value of the ratio \Mdot$/\sqrt{f_{\rm cl}(r)}$ as a function of $r/R_{\star}$ 
for a typical O type star (relevant parameters are color-coded and listed above and in the body 
of each panel). This ratio controls the wind emission due to recombination of ionized hydrogen, 
i.e., the higher its value, the greater the emission. We used Eq.~(\ref{e:clumping}) with 
the different values of \Vcl\ (color-coded in the body of each figure) to calculate $f_{\rm cl}$.}
\label{f:clumping}
\end{figure*}

Fig.~\ref{f:A2} and also Fig.~\ref{f:H_best} illustrate how difficult it is to estimate the parameters \Fcl\ and $\beta$. We have essentially one diagnostic, H$\alpha$, which is affected by multiple other parameters. 
Fitting H$\alpha$ is also where our results differ the most from those of \cite{pue16}. They found 
that the highly clumped wind (\Fcl$<0.05$) fits this line the best, while our results suggest a much 
lower degree of clumping (\Fcl$\sim 0.6$) and a higher mass-loss rate. Most of the difference originates 
from the different radial distribution of the true \textit{\textup{volume filling factor}}, $f_{\rm cl}(r)$, in 
our models (see Eq.~(\ref{e:clumping}) for its definition). While we use a generic value \Vcl$=0.1$\Vinf\ 
in Eq.~(\ref{e:clumping}), they used a very low value of \Vcl$\sim 50$~km s$^{-1}$. The value we use 
creates a large transitional zone at the base of the wind for slowly accelerating models ($\beta >1.7$), 
while their models always reach the prescribed volume filling factor very rapidly.  \Fcl\ is 
the volume filling factor at large radii, where $v(r)\sim$\Vinf\ in the parameterization of CMFGEN, and 
it does not mean that $f_{\rm cl}(r)$=\Fcl\ at every radius. In Fig.~\ref{f:clumping} we show the true 
radial distribution of \Mdot$/\sqrt{f_{\rm cl}(r)}$ calculated by using our best-fitting $\beta$ and \Vcl\ parameters for a typical O star, and what these models 
would have if we had used the \Vcl\ value of \cite{pue16}. The quantity \Mdot$/\sqrt{f_{\rm cl}(r)}$ is 
important because it controls the emission by the recombination of \ion{H}{II} in the wind (assuming 
that the ionization structure of hydrogen is similar in the models). The panels in Fig.~\ref{f:clumping} show 
that the values of \Mdot$/\sqrt{f_{\rm cl}(r)}$ are always higher at every radius for models with 
smaller \Vcl\ (and with all other parameters being the same), and this excess is greater and extends to 
much larger radii for highly clumped (low \Fcl) and slowly accelerating (high $\beta$) models. 
This means that the models of \cite{pue16} produce much more emission in the crucial dense internal part of 
the wind than ours, especially for smaller \Fcl\ and larger $\beta$. This explains why our results 
favor a higher mass-loss rate and a lower level of clumping when the same H$\alpha$ profile is analyzed. 
Finally, we would like to stress that the difference would be smaller if the wind of $\epsilon$ Ori had 
accelerated normally ($\beta <1$).

It is difficult to judge which distribution is more realistic because the hydrodynamic simulations are 
quite fuzzy on this subject \citep[see, e.g.,][]{run02}. The simulations suggest that the wind is 
smooth near the stellar surface and predict a transitional zone, but they are not clear about the 
size of this zone. The motivation behind using an Eq.~(\ref{e:clumping}) type distribution is exactly 
to reproduce these characteristics and to recognize that the clumping scales with the wind velocity, 
that is, the faster the wind, the more clumped. The wind of $\epsilon$~Ori is already anomalous in 
the sense that it likely accelerates slowly. Theoretical calculations \citep[e.g.,][]{CAK} predict 
$\beta$-values lower than unity. The question then is that if a star has a slowly accelerating 
wind ($\beta\sim 2$) and if the clumping scales with the wind velocity, why would the prescribed \Fcl\ at 
\Vinf\ be reached rapidly? Nevertheless, we are not in the position to decide which distribution 
is more realistic, therefore we consider all the models with \Fcl$=0.05$, \Fcl$=0.3$, and \Fcl$=0.6$ as 
best-fitting models. However, we note that if our finding of a higher mass-loss rate is correct, it indirectly supports the idea that the UV resonance lines are strongly affected by 
vorosity. 

Finally, we would like to comment on the evolutionary status of $\epsilon$~Ori. A pleasant side-effect of 
using actual evolutionary calculations to create stellar atmosphere models is that the result comes with immediate information on the evolutionary status of the subject star.  As shown in the H-R diagrams of Figs.~\ref{f:HRDs_0.05nrot}--\ref{f:HRDs_0.6nrot} as well as in Table~\ref{tab:parameters}, the locations of
the best-fitting models are near the terminal age  main sequence (TAMS) of the evolutionary track for M$_{\rm ini} \sim$ 40\Msun. Furthermore, Table~\ref{tab:parameters} shows that all the best-fitting models are from the batch that was calculated without considering the rotational enhancement in the surface abundances, which indirectly suggest that the star rotates slowly. This is consistent with the fact $\epsilon$~Ori has low \vsini.  

\section{Summary}
\label{sec:concl}
We presented a mega grid of 43,340 stellar atmospheric models calculated by the CMFGEN package, which
will soon be extended to 80,000 models. These models cover the region of the H--R diagram that is populated by 
OB main-sequence and W-R stars with masses between 9 and 120 $M_{\odot}$. The grid provides UV, 
visual, and IR spectra for each model. 

We used the surface temperature (\Teff) and luminosity ($L$) values that correspond to the
evolutionary traces and isochrones of \cite{eks12}. Furthermore, we used seven values of $\beta$, four values of 
the clumping factor, and two different metallicities and terminal velocities. This generated a 
six-dimensional hyper-cube of stellar atmospheric models that we intend to release to the 
general astronomical community as a free tool for analyzing the spectra of massive stars.

We have also demonstrated the usefulness of our mega-grid by reanalyzing $\epsilon$~Ori.
Our somewhat crude but very rapid analysis supported the stellar and wind parameters reported by 
\cite{pue16}. The only significant difference is that our reanalysis favors the high end of the 
acceptable mass-loss range (\Mdot$\sim 1.1\times 10^{-6}$ $M_{\odot}$ yr$^{-1}$) with a lower level 
of clumping. This result indirectly supports recent simulations that suggest that the UV resonance 
lines are highly affected by velocity-space porosity. The reason for the slightly different conclusion 
of our reanalysis is that we used a generic radial distribution of clumping, while \cite{pue16} 
customized the distribution to $\epsilon$ Ori. It is not clear that our distribution
is worse.

The reanalysis showed the benefits of having a large grid of precalculated models. The stellar and wind parameters for a star can be calculated rapidly. If required, a more detailed study can then be 
performed, but by starting with good initial values. This significantly shortens the time that is 
needed to complete a spectral analysis.    

\begin{acknowledgements}
All models and their synthetic spectra were calculated by the cluster
Abacus I. The authors express their acknowledgement for the resources,
expertise and the assistance provided by "ABACUS" Laboratory of
Applied Mathematics and High Performance Computing CINVESTAV-IPN,
CONACyT-EDOMEX-2011-C01-165873 Project.  The authors are also grateful to 
D.J. Hillier, the author of the  code CMFGEN, for his helpful comments during the production of the 
grid and during the preparation of this article. J. Zsargo acknowledges
CONACyT CB-2011-01 No. 168632 grant for support. J. Klapp acknowledges
financial support by the Consejo Nacional de Ciencia y Tecnolog\'ia
(CONACyT) of M\'exico under grant 283151. The authors also acknowledge the anonymous 
referee for his or her helpful comments and suggestions.
\end{acknowledgements}

\bibliographystyle{aa}
\bibliography{manuscript38066.bbl}
 
\begin{appendix}

\section{}

\begin{figure*}
\centering
\includegraphics[width=0.8\textwidth]{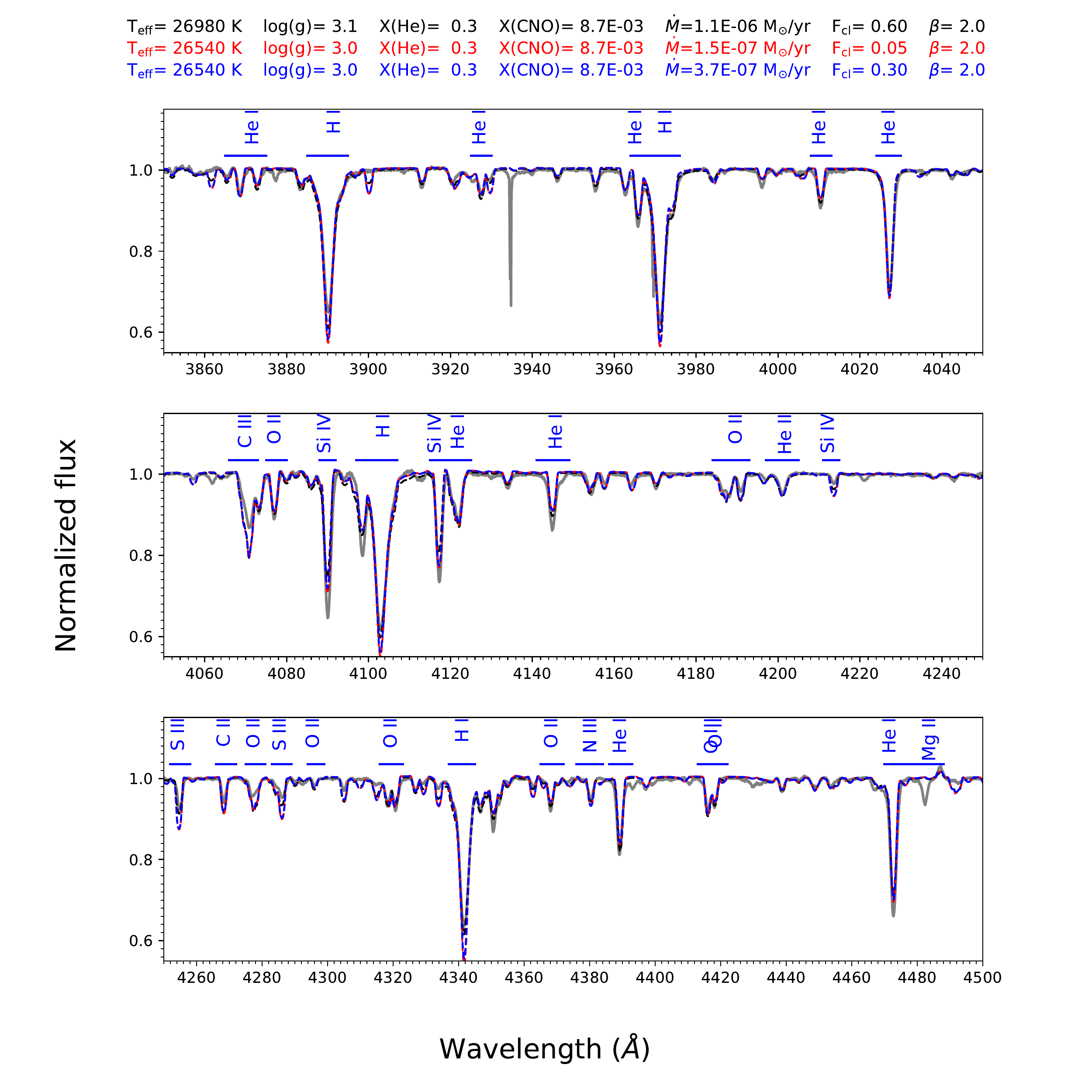}
\caption{Comparison of our best-fitting models (dashed black, red, and blue lines) with the blue part of the optical spectra for $\epsilon$~Ori (solid gray line). The relevant parameters are color-coded and listed above the figure, and important lines are indicated above the spectra.}
\label{f:A1}
\end{figure*}

\begin{figure*}
\centering
\includegraphics[width=0.8\textwidth]{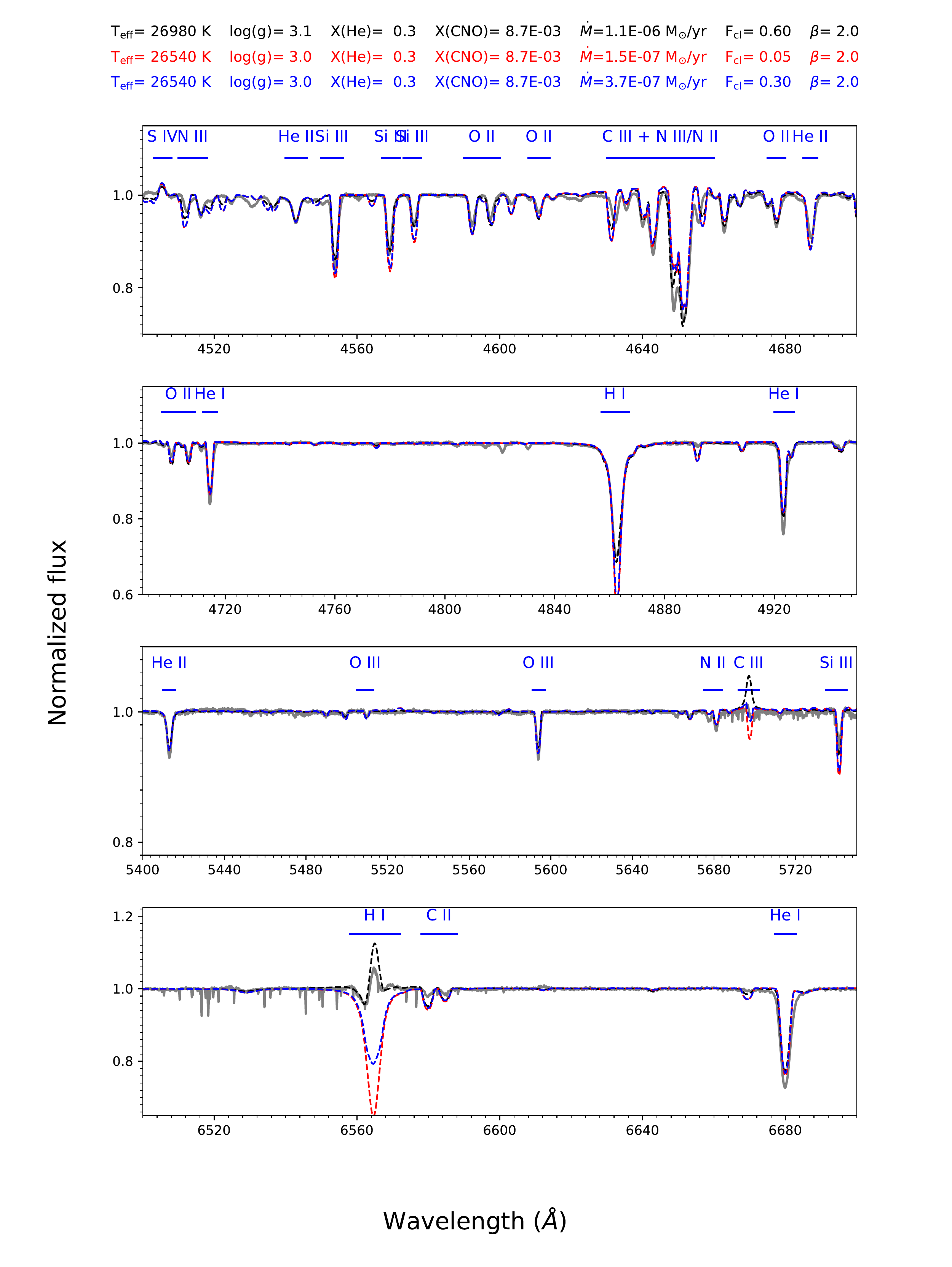}
\caption{Same as Fig.~\ref{f:A1}, but for the central region of the optical spectra.}
\label{f:A2}
\end{figure*}


\begin{figure*}
\centering
\includegraphics[width=0.34\textwidth]{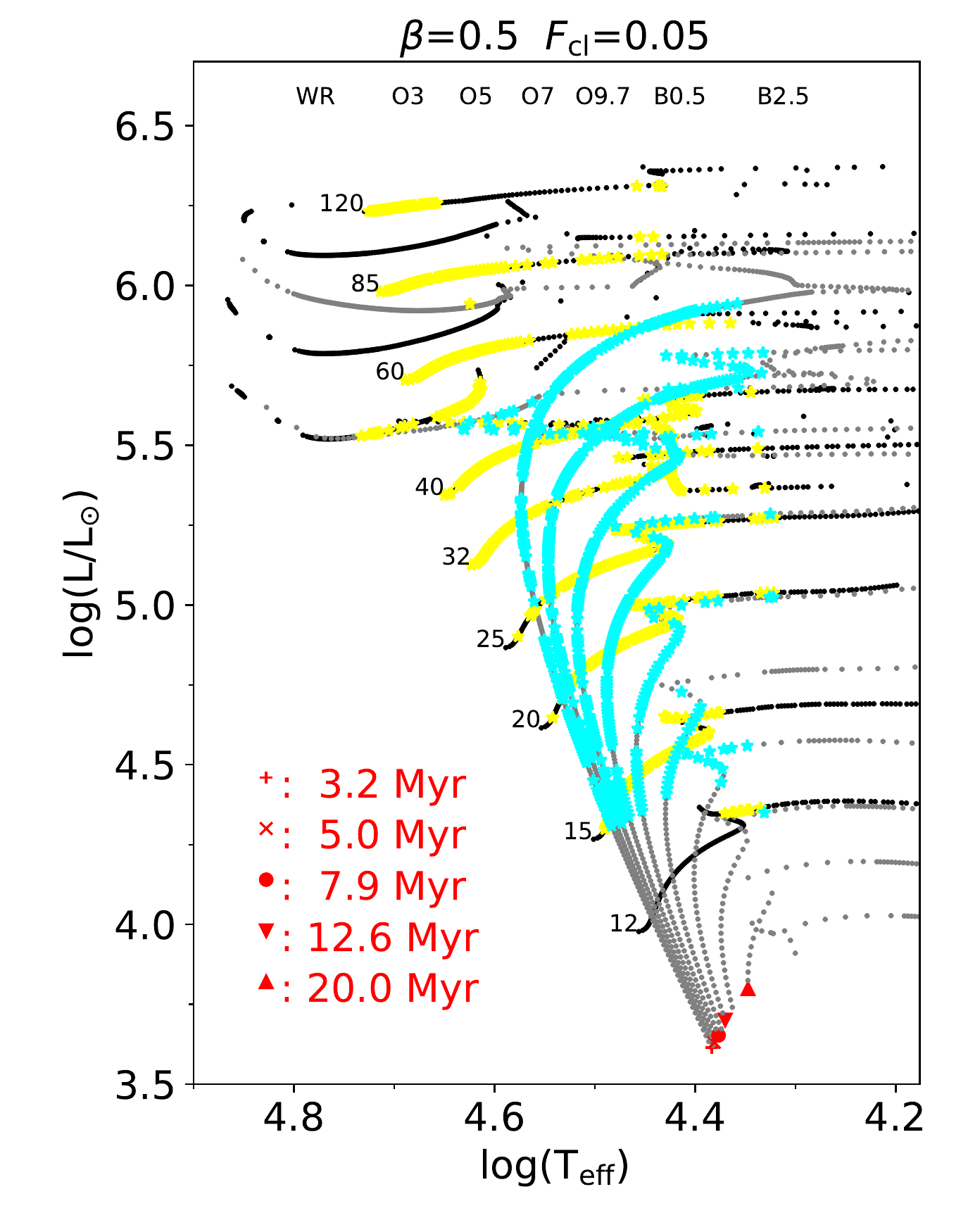}
\includegraphics[width=0.34\textwidth]{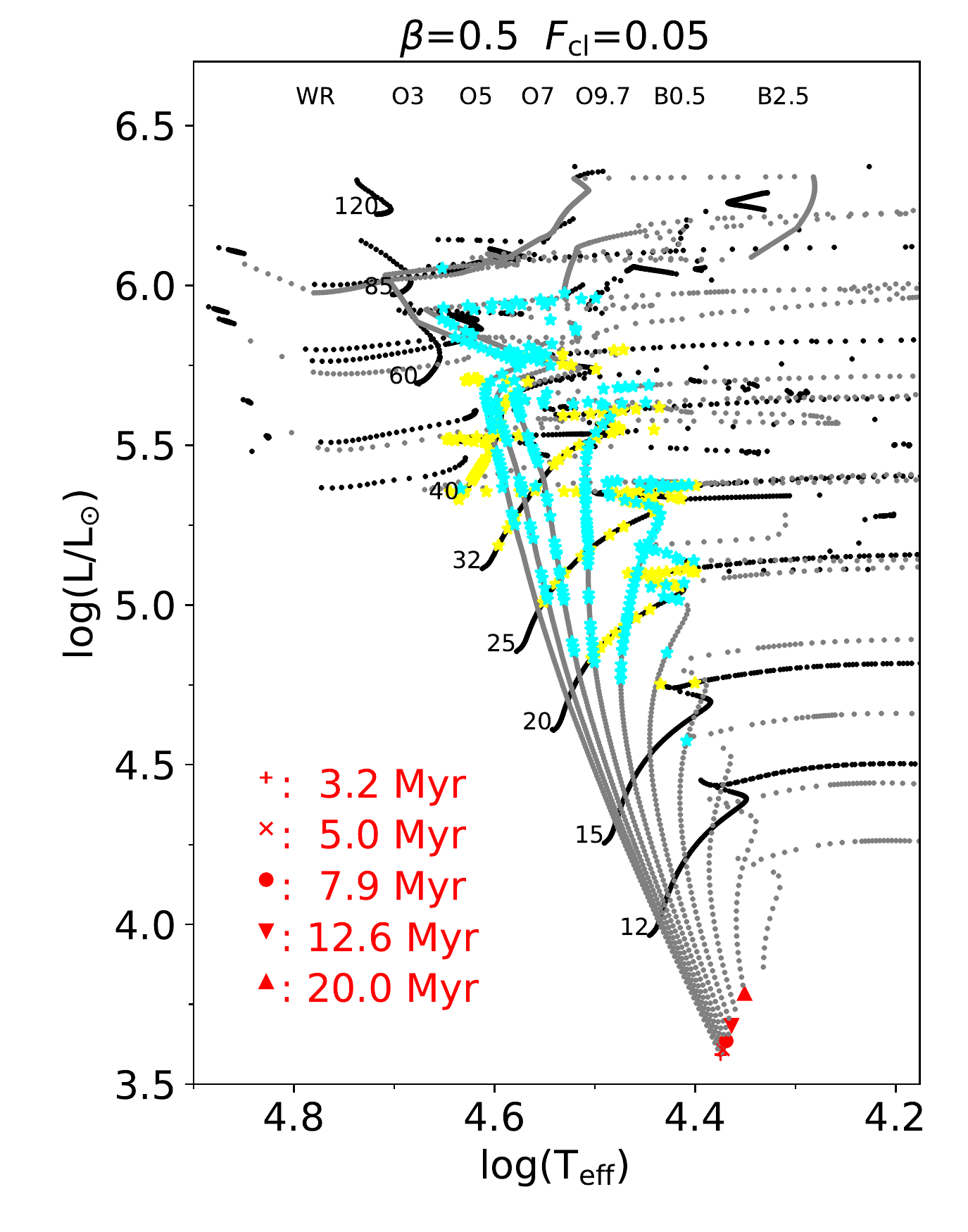}
\includegraphics[width=0.34\textwidth]{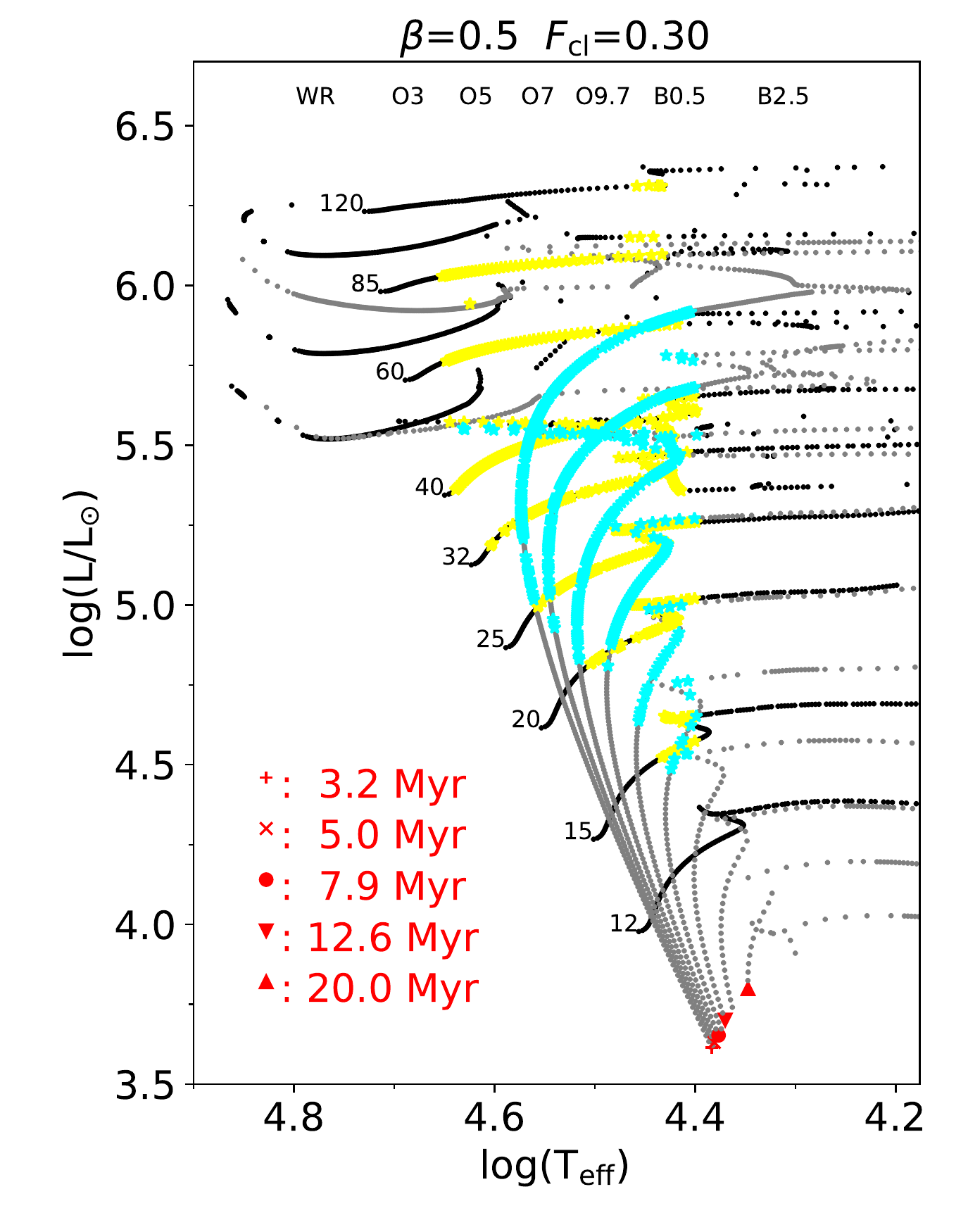}
\includegraphics[width=0.34\textwidth]{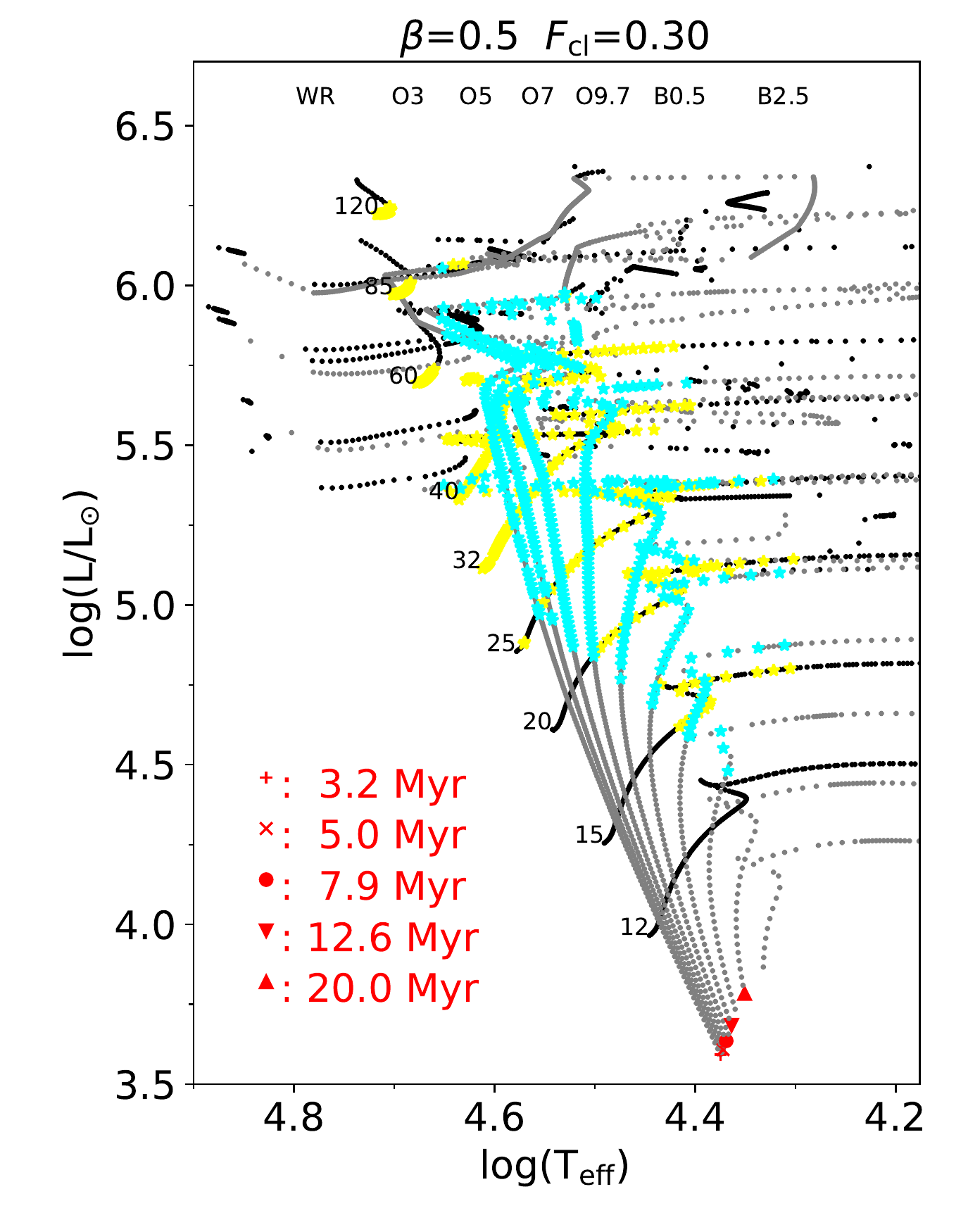}
\includegraphics[width=0.34\textwidth]{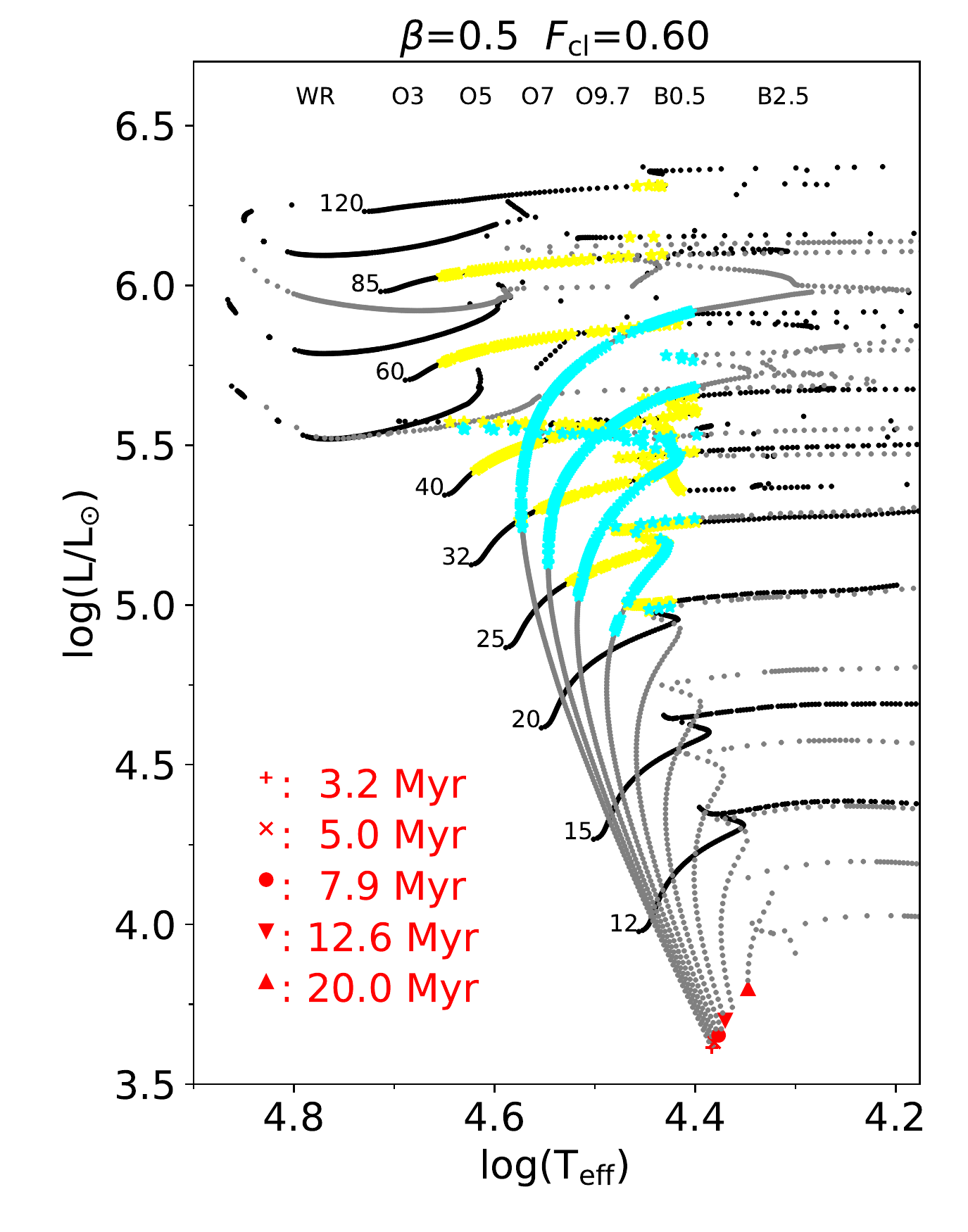}
\includegraphics[width=0.34\textwidth]{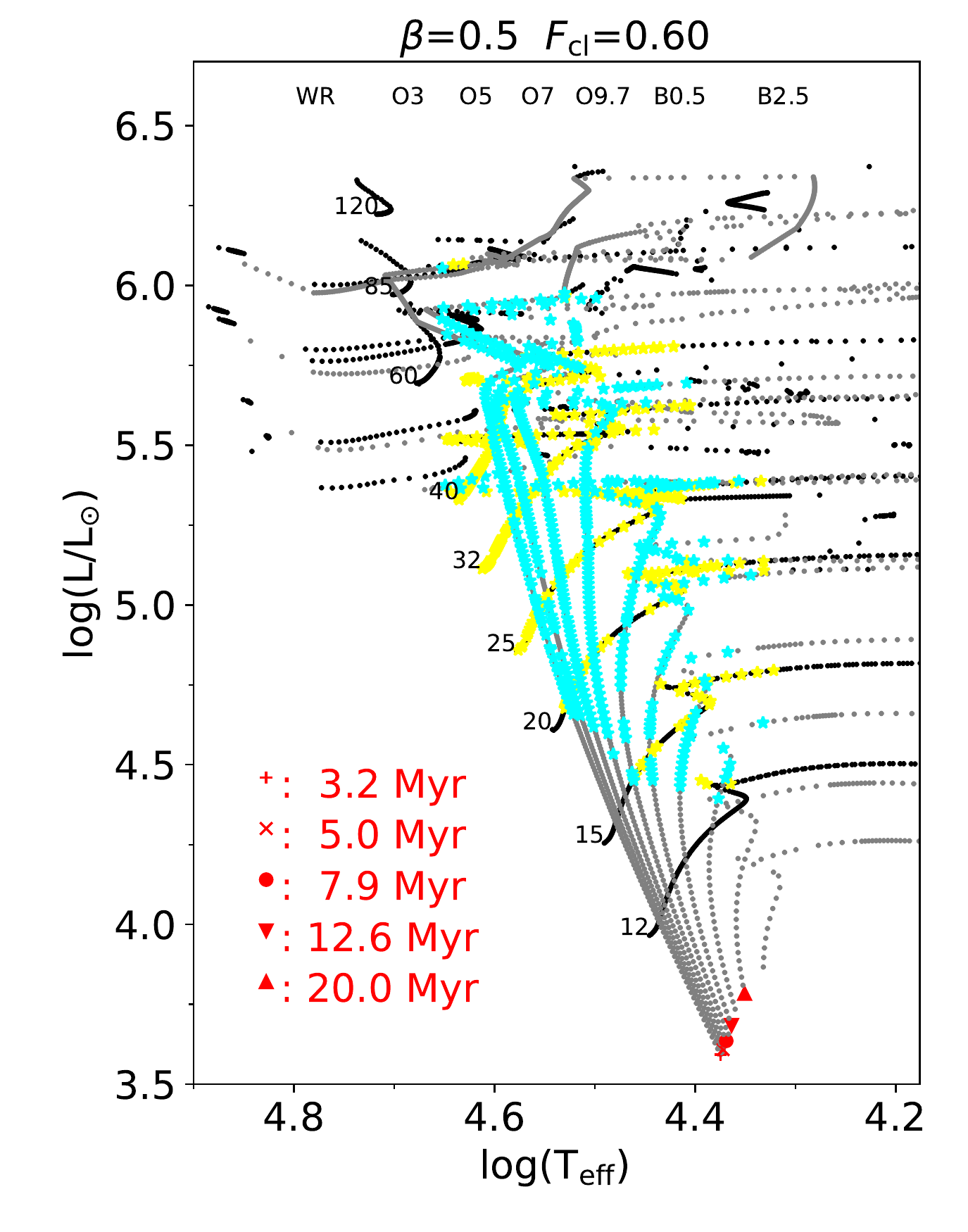}
\caption{Sample of H-R diagrams, similar to Fig.~\ref{f:HRDs_0.05nrot}, for various \Fcl volume filling factors. The left and right columns show models without and with rotational enhancement, respectively.}
\label{f:HRDs_appx}
\end{figure*}

\end{appendix}

\label{lastpage}
\end{document}